\documentclass[aps, prd, twocolumn, lengthcheck, superscriptaddress, showpacs, letterpaper, nofootinbib]{revtex4-1}
\usepackage{amsfonts}
\usepackage{amsmath}
\usepackage{amssymb}
\usepackage{xcolor}
\usepackage{graphicx}
\usepackage{longtable}
\usepackage{slashed}
\usepackage{hyperref}
\usepackage{amsthm}
\usepackage{mathrsfs}
\usepackage{color}

\def\be{\begin{eqnarray}}\def\ee{\end{eqnarray}}

\allowdisplaybreaks

\begin{document}
\title{Chiral effective Lagrangian for heavy-light mesons from QCD: $1/m_{Q}$ correction}

\author{Qing-Sen Chen}
\affiliation{Center for Theoretical Physics, College of Physics, Jilin University,
Changchun 130012, China}

\author{Hui-Feng Fu}
\email{huifengfu@jlu.edu.cn}
\affiliation{Center for Theoretical Physics, College of Physics, Jilin University,
Changchun 130012, China}

\author{Yong-Liang Ma}
\email{ylma@ucas.ac.cn}
\affiliation{School of Fundamental Physics and Mathematical Sciences,
Hangzhou Institute for Advanced Study, UCAS, Hangzhou, 310024, China}
\affiliation{International Center for Theoretical Physics Asia-Pacific (ICTP-AP) (Beijing/Hangzhou), UCAS, Beijing 100190, China}

\author{Qing Wang}
\email{wangq@mail.tsinghua.edu.cn}
\affiliation{Department of Physics, Tsinghua University, Beijing 100084, China
}
\affiliation{Center for High Energy Physics, Tsinghua University, Beijing 100084, China
}

%

\begin{abstract}
As a successive work to [Phys.Rev.D 102 (2020), 034034], we derive the $1/m_Q$ corrections to chiral effective Lagrangian for heavy-light mesons from QCD under proper approximations. The low energy constants in the effective Lagrangian are expressed in terms of the light quark self-energy and heavy quark mass $m_Q$. Numerical results of the low energy constants with $1/m_Q$ corrections are given. We find that the results of pion decay constant and the masses of heavy-light mesons are improved coherently compared to that obtained in the heavy quark limit.
\end{abstract}


 \maketitle

\section{Introduction}
Establishing the analytic relationships between the low energy constants (LECs) of the chiral effective theory and QCD has great significance in hadron physics.
Such relationships fill up the gap between the effective theories and the fundamental theory, and allow one to calculate the LECs from the QCD Green functions. This type of researches has been carried out in Refs.~\cite{Wang:1999cp,Yang:2002hea,Jiang:2009uf,Jiang:2010wa,Jiang:2012ir,Jiang:2015dba} for the traditional chiral effective theory and in Refs.~\cite{Wang:2000mu,Wang:2000mg,Ren:2017bhd} for some extensions to the chiral effective theory. Recently, the chiral effective theory for heavy-light mesons were derived from QCD in the heavy quark limit $m_Q\rightarrow \infty$~\cite{Chen2020jiq}.

In Ref.~\cite{Chen2020jiq}, we focused on the heavy-light meson doublets with the spin-parity of the light quark  cloud $s_l^P=\frac{1}{2}^-$ (or $J^P=(0^-, 1^-)$ for the spin-parity of the heavy-light mesons) and $s_l^P=\frac{1}{2}^+$ (or $J^P = (0^+, 1^+)$) which are denoted as $H$ and $G$ respectively. The LECs in the effective Lagrangian (including the masses of the mesons and the coupling constants) were expressed in terms of the light quark self-energy which can be calculated using Dyson-Schwinger equations or lattice QCD. The resulted numerical values turned out to be roughly consistent with the experimental data. The mass splitting of $H$ and $G$  doublets emerges evidently from dynamical chiral symmetry breaking of QCD which is realized by the solution to the quark gap equation.

Soon after the early works on the chiral effective theory for heavy-light mesons in the heavy quark limit~\cite{Wise:1992hn,Yan:1992gz,Nowak:1992um,Casalbuoni:1992dx}, the effects due to finite heavy quark mass began to be explored~\cite{Cheng:1993gc,Balk:1993ev,DiBartolomeo:1994ir} and continue to be an important topic until now~\cite{Cheung:2015rya,Alhakami:2019ait}. Since our framework establishes the relationships between the LECs and the fundamental theory QCD, it provides a method to calculate the $1/m_Q$ corrections from QCD. We study in this work the $1/m_Q$ corrections to the LECs of the chiral Lagrangian for the heavy-light mesons in order to improve our previous results. It is found that the $1/m_Q$ corrections are indeed helpful in improving the numerical results. Especially, the tension between the pion decay constant $f_\pi$ and the mass splitting, when calculated up to the $1/m_Q$ order, is released. We study both the charmed mesons and the bottom mesons in this paper, for these two meson sectors suffer from different values of $1/m_Q$ corrections, which provides more detailed information on the validity of the heavy quark expansion.

This paper is organized as follows. In Sec.~\ref{sec:HeavyEFT}, we introduce the chiral effective Lagrangian for the heavy-light mesons. In Sec.~\ref{sec:EFTQCD}, the LECs of the chiral Lagrangian up to $1/m_Q$ order is derived from QCD, which turn to be integrals of the dressing functions of the light quark propagator. The numerical results based on these formula are given in Sec.~\ref{sec:Num}, where the relevant dressing functions of the quark propagator are obtained from the quark Dyson-Schwinger equation as well as from lattice QCD. A summary is given in Section~\ref{sec:dis}.

\section{chiral effective Lagrangian for heavy-light mesons}
\label{sec:HeavyEFT}

For convenience, we introduce the chiral effective Lagrangian for heavy-light mesons in this section. The heavy-light meson doublets $H$ and $G$ can be expressed as
\begin{eqnarray}
H & = & \frac{1 + v\hspace{-0.15cm}\slash}{2}\left(P^{\ast;\mu}\gamma_\mu + i P \gamma_5\right),\nonumber\\
G & = & \frac{1 + v\hspace{-0.15cm}\slash}{2}\left(S^{\ast;\mu}\gamma_\mu\gamma_5 + S\right),
\label{eq:HeavyMeson}
\end{eqnarray}
where $(P,P^{\ast;\mu})$ refer to $J^P=(0^-, 1^-)$ states, and $(S,S^{\ast;\mu})$ refer to $J^P = (0^+, 1^+)$ states, respectively. $v^\mu$ is the velocity of an on-shell heavy quark, i.e. $p^\mu=m_Q v^\mu$ with $v^2=1$.
As in Ref.~\cite{Chen2020jiq}, we only consider two light flavors. Then the chiral effective Lagrangian for heavy-light mesons (up to first order derivatives) can be written as~\cite{Manohar:2000dt,Nowak:2004jg}
\begin{eqnarray}
{\cal L} & = & {\cal L}_H + {\cal L}_G + {\cal L}_{HG},
\label{eq:LHHChPT}
\end{eqnarray}
where~\footnote{Here, since we are not interested in the mass splitting of the hadrons in a heavy quark doublet, we shall not write done the effective Lagrangian containing magnetic moment operator. }
\begin{eqnarray}
{\cal L}_H & = & {}-i{\rm Tr} \left(\bar{H} v\cdot \nabla H\right) - g_H {\rm Tr} \left(H \gamma^\mu\gamma_5 A_\mu  \bar{H}\right) \nonumber\\
& &{} + m_H{\rm Tr} \left(\bar{H} H\right),\nonumber\\
{\cal L}_G & = &{} - i {\rm Tr} \left(\bar{G} v\cdot \nabla G\right) + g_G {\rm Tr} \left(G \gamma^\mu\gamma_5 A_\mu  \bar{G} \right) \nonumber\\
& &{} + m_G{\rm Tr} \left(\bar{G} G\right), \nonumber\\
{\cal L}_{HG} & = & g_{HG}{\rm Tr} \left(H \gamma^\mu\gamma_5 A_\mu  \bar{G}\right) + h.c.,
\end{eqnarray}
where the covariant derivative $\nabla_\mu = \partial_\mu - i V_\mu$ with $V_\mu= \frac{i}{2}\left(\Omega^\dagger \partial_\mu \Omega + \Omega \partial_\mu \Omega^\dagger\right)$, and $A_\mu = \frac{i}{2}\left(\Omega^\dagger \partial_\mu \Omega - \Omega \partial_\mu \Omega^\dagger\right)$. The $\Omega$ field is related to the chiral field $U(x) = \exp(i\pi(x)/f_\pi)$ through $U = \Omega^2$.  The parameters $g_H, g_G, g_{HG}, m_H$, and $m_G$ are the LECs of the Lagrangian. They are free ones at the level of effective theory since they cannot be controlled by symmetry argument which the effective theory relies on.

The states associated with $H$ and $G$ are called chiral partners~\cite{Nowak:1992um,Bardeen:1993ae}, and their mass splitting arises from the chiral symmetry breaking. In the $D$ meson family, where $H$ is associated to $(D, D^*)$ and $G$ is associated to $(D^*_0, D_1)$, the spin-averaged masses of the chiral partners are~\cite{ZylaPDG}
\begin{eqnarray}\label{eq:EXmass}
m_H & \simeq & 1970~{\rm MeV}, \;\;\; m_G \simeq 2400~{\rm MeV},
\end{eqnarray}
which yields the mass splitting
\begin{eqnarray}
\Delta m & = & m_G - m_H = 430~{\rm MeV}.
\label{eq:splitting}
\end{eqnarray}
In the $B$ meson family, where $H$ is associated to $(B, B^*)$ and $G$ is associated to $(B^*_0, B_1)$, the spin-averaged masses of the chiral partners are~\cite{ZylaPDG}
\begin{eqnarray}\label{eq:EXmass'}
m_H & \simeq & 5302~{\rm MeV}, \;\;\; m_G \simeq 5726~{\rm MeV},
\end{eqnarray}
and the mass splitting is
\begin{eqnarray}
\Delta m & = & m_G- m_H = 424~{\rm MeV}.
\label{eq:splitting'}
\end{eqnarray}

\section{ The $1/m_{Q}$ corrections to chiral effective Lagrangian for heavy-light mesons from QCD}
\label{sec:EFTQCD}

To obtain the $1/m_{Q}$ corrections to the chiral effective Lagrangian for the heavy-light mesons, we follow our previous analysis but retain only the $1/m_Q$ order contributions in the heavy quark expansion. The QCD generating functional can be written as
\begin{widetext}
\begin{eqnarray}
 	Z[J] & = &\int \mathcal{D}q\mathcal{D} \bar q \mathcal{D}Q \mathcal{D} \bar Q  \mathcal{D} G_{\mu}\varDelta_F(G_{\mu}) \exp \left\{ i\int d^4x \left[ \bar q (i\slashed{D}+J)q+ \bar Q (i\slashed{D}-m_Q)Q  -\frac{1}{4}G_{\mu\nu}^iG^{\mu\nu}_i-\dfrac{1}{2\xi}[F^i(G_{\mu})]^2\right]  \right\},
 	\label{generating1}
\end{eqnarray}
\end{widetext}
where $q(x)$, $Q(x)$ and $G_\mu(x)$ are the light-quark, heavy-quark and gluon fields, respectively. $J(x)$ is an external source for the composite light quark fields. The masses of the light quarks are absorbed in the external source $J(x)$ and would vanish in the chiral limit. The heavy quark mass is denoted as $m_Q$. The indices $i, j, \cdots$  represent the color indices in the adjoint representation.

Following the standard heavy quark effective theory (HQET)~\cite{Manohar:2000dt}, we formulate the effective Lagrangian directly in terms of the velocity-dependent fields $N_v(x)$ and $\mathcal{N}_v(x)$:
\begin{eqnarray}
N_v(x) & = & \frac{1+\slashed{v}}{2}e^{im_Q v\cdot x} Q(x), \nonumber\\
\mathcal{N}_v(x) & = & \frac{1-\slashed{v}}{2}e^{im_Q v\cdot x} Q(x),
\label{heavy form}
\end{eqnarray}
where $N_v(x)$ is the large component of the heavy quark field and $\mathcal{N}_v(x)$ is the component of the heavy quark field suppressed by power of $1/m_Q$. The factor $e^{im_Q v\cdot x}$ rotates away the mass of the heavy quark field.
Integrating out $\mathcal{N}_v(x)$ at tree level amounts to the replacement
\begin{eqnarray}
\mathcal{N}_v(x) = \frac{i\slashed D_{\perp} }{i v\cdot D +2m_Q} N_v,
\label{Nrelation}
\end{eqnarray}
where $ \slashed D_{\perp}=\slashed D-  v\cdot D \slashed v  $, is the perpendicular component of $\slashed D$. Then the heavy quark part of the Lagrangian $\mathcal{L}_Q=\bar Q (i\slashed{D}-m_Q) Q$ becomes
\begin{equation}\label{HQ}
\begin{array}{ccl}
\mathcal{L}_Q &=& \bar N_v \left ( i v\cdot D -\frac{\slashed D_{\perp} \slashed D_{\perp} }{i v\cdot D +2m_Q}  \right )  N_v \\
&=&  \bar N_v \left ( i v\cdot \partial -\frac{(\slashed \partial_{\perp})^2 }{2m_Q} \right ) N_v +\mathcal{L}_{int}^{N_vG},
\end{array}
\end{equation}
where $ (\slashed \partial_{\perp})^2=(\slashed \partial-  v\cdot \partial \slashed v)^2 = \partial^2-(v\cdot\partial)^2 $ and $\mathcal{L}_{int}^{N_vG}$ accounts for the interaction between the heavy quark field $N_v$ and the gluon field. Up to the $1/m_Q$ order, the generating functional becomes
\begin{widetext}
\begin{eqnarray}
Z[J] & = &\int \mathcal{D}q\mathcal{D} \bar q \mathcal{D} N_v \mathcal{D} \bar N_v \exp \left\{ i\int d^4x \left[ \bar q (i\slashed{\partial}+J)q+ \bar N_v \left ( i v\cdot \partial -\frac{(\slashed \partial_{\perp})^2 }{2m_Q} \right ) N_v \right]\right\}  \notag \\
& &{}~~\times\int \mathcal{D} G_{\mu}\varDelta_F(G_{\mu})\exp\left\{i \int d^4x\left[{}-\frac{1}{4}G_{\mu\nu}^iG^{\mu\nu}_i-\dfrac{1}{2\xi}[F^i(G_{\mu})]^2-g \mathcal{I}^{\mu}_i G^i_{\mu}  +\mathcal{L}_{int}^{N_vG}  \right] \right\},
\label{generating2}
\end{eqnarray}
\end{widetext}
where $\mathcal{I}_i^{\mu}=\bar q \frac{\lambda_i}{2}\gamma^{\mu} q $ is the light quark current.

The chiral effective action for the heavy-light mesons can be obtained by first integrating in the chiral field $U(x)$ and the heavy-light meson fields $\sim (q\bar{N}_v)$ to and then integrating out gluon fields and quark fields from the generating functional~\eqref{generating2}. Taking the large $N_c$ limit and keeping the leading order in the dynamical perturbation, we obtain the effective action as
\begin{eqnarray}\label{action2}
S[U,\Pi_2,\bar{\Pi}_2] & = &-i N_c \mathrm{Tr}^\prime \ln\Bigg[(i\slashed{\partial}-\bar{\Sigma})I_1 +J'_\Omega \nonumber\\
&&{}\qquad\qquad\quad\, + \left(i v\cdot \partial -\frac{(\slashed \partial_{\perp})^2 }{2m_Q} \right)I_4\nonumber\\
&&{}\qquad\qquad\quad\, -\Pi_2-\bar{\Pi}_2 \Bigg],
\end{eqnarray}
where $\Pi_2$ and $\bar{\Pi}_2$ are the heavy-light meson field and its conjugate, respectively. $\bar{\Sigma}$ is the self-energy of the light quark propagator. $J'_\Omega$ is the chiral-rotated external source which the chiral field is attached on. $I_1=\mathrm{diag}(1,1,0)$ and $I_4=\mathrm{diag}(0,0,1)$ are the light- and heavy- projecting matrices in the flavor space respectively. $\mathrm{Tr}^\prime$ represents a functional trace over the flavor space, spinor space, and coordinate space. The details of the derivation of Eq. (\ref{action2}) are given in Appendix \ref{sec:AppA}.

Since our purpose is to obtain the chiral effective Lagrangian for the chiral partners given by Eq.~\eqref{eq:HeavyMeson}, we only keep the $H$ and $G$ fields in $\Pi_2$:
\begin{eqnarray}
\Pi_{2}(x) & = & H^q(x)+G^q(x), 
\end{eqnarray}
where $q=1, 2$ is light flavor indices. The chiral effective Lagrangian is generated by expanding the action $S[U,\Pi_2,\bar{\Pi}_2]$ with respect to the fields $U$, $\Pi_2$ and $\bar{\Pi}_2$. The kinetic terms of the heavy-light meson fields arise from
\begin{eqnarray}
S_{2} & \equiv & \frac{\delta^2 S}{\delta \bar{\Pi}_2 \delta \Pi_2} \Pi_2\bar{\Pi}_2\nonumber\\
& = & iN_c\int d^4x_1 d^4 x_2 \nonumber\\
& & {} \times \mathrm{Tr}\Bigg[\left(i\slashed{\partial} -\bar{\Sigma}\right)^{-1}\delta(x_2-x_1)\bar{H}(x_1)\nonumber\\
&&\qquad\qquad \times\left(i v\cdot\partial-\frac{(\slashed \partial_{\perp})^2 }{2m_Q}\right)^{-1}\delta(x_1-x_2) H(x_2)\Bigg],\nonumber\\
\label{S2}
\end{eqnarray}
where $\bar{\Sigma}(x-y)=\Sigma(\nabla^2)\delta(x-y)$ with $\Sigma$ being the light-quark self-energy function in the coordinate space. We take the argument of $\Sigma$ to be the covariant derivative in order to retain the correct chiral transformation properties in the theory~\cite{Yang:2002hea}. We have considered only the field $H$ in Eq. (\ref{S2}), while a similar equation holds for $G$. Taking the derivative expansion up to the first order, we obtain
\begin{widetext}
\begin{eqnarray}
S_2 & = & S^{(0)}_{2}+S^{(1)}_{2}\notag\\
S^{(0)}_{2} & = & i N_c\int d^4x \int\frac{d^4  p}{(2\pi)^4} \left[-\frac{1}{ p^2-\Sigma^2} +\frac{\Sigma}{( p^2-\Sigma^2)v\cdot  p}\right] \mathrm{Tr}\left[\bar H(x)H(x)\right]\notag\\
&&{} +iN_c\int d^4x \int\frac{d^4  p}{(2\pi)^4} \left[-\frac{1}{( p^2-\Sigma^2) v\cdot  p}- \frac{2\Sigma}{( p^2-\Sigma^2)^2}-\frac{2\Sigma'( p^2+\Sigma^2)}{( p^2-\Sigma^2)^2}\right]\mathrm{Tr}\left[ H(x)iv\cdot\partial \bar H(x)\right]\notag\\
&&{} +iN_c\int d^4x \int\frac{d^4  p}{(2\pi)^4} \left[-\frac{2\Sigma'( p^2+\Sigma^2)}{( p^2-\Sigma^2)^2}\right]\mathrm{Tr}\left[ H(x)v\cdot V_\Omega \bar H(x)\right], \notag \\
S^{(1)}_{2} & = & i N_c\int d^4x \int\frac{d^4  p}{(2\pi)^4} \left[-\frac{1}{ p^2-\Sigma^2} +\frac{\Sigma}{( p^2-\Sigma^2)v\cdot  p}\right]\left[\frac{-p^2+(v \cdot p)^2}{2m_Q}\right] \mathrm{Tr}\left[\bar H(x)H(x)\right]\notag\\
&&{} +iN_c\int d^4x \int\frac{d^4  p}{(2\pi)^4} \biggr[{} -\frac{1}{(p^2-\Sigma^2) v\cdot p} - \frac{2\Sigma+2\Sigma'( p^2+\Sigma^2)- (4\Sigma\Sigma'+2)v\cdot p}{(p^2-\Sigma^2)^2}\biggr] \biggr[\frac{-p^2+ (v\cdot p)^2}{2 m_Q v\cdot p }\biggr]\nonumber\\
& &\qquad\qquad\qquad\qquad\quad {} \times \mathrm{Tr}\left[ H(x)iv\cdot\partial \bar H(x)\right]\notag\\
&&{} +iN_c\int d^4x \int\frac{d^4  p}{(2\pi)^4} \left[-\frac{2\Sigma'( p^2+\Sigma^2)-4\Sigma\Sigma'v\cdot p}{( p^2-\Sigma^2)^2}\right]\biggr[\frac{-p^2+ (v\cdot p)^2}{2 m_Q v\cdot p }\biggr] \mathrm{Tr}\left[ H(x)v\cdot V_\Omega \bar H(x)\right],
	\label{eq:S22}
\end{eqnarray}
\end{widetext}
$S^{(0)}_{2}$ is the leading order in the $1/m_Q$ expansion (which survives in the $m_Q\rightarrow \infty$ limit) and $S^{(1)}_{2} $ is the $1/m_Q$ order correction to $S^{(0)}_{2}$. The simplest terms of the heavy-light mesons interacting with the Goldstone boson are generated by taking additional derivative with respect to the chiral field $U(x)$. Since the $U$ field is attached to the rotated external source, we actually take derivatives on the action $S$ with respect to $J_\Omega$. The resultant expression is
\begin{eqnarray}\label{S3}
S_{3} & \equiv & \frac{\delta^3 S}{\delta J_\Omega  \delta \bar H \delta  H } J_\Omega H\bar{H},\notag\\
& = &{} -iN_c\int d^4x_1d^4x_2d^4x_3\nonumber\\
&&\qquad {} \times \mathrm{Tr}\Bigg[(i\slashed{\partial} -\Sigma)^{-1}\delta (x_2-x_3)\nonumber\\
&&\qquad\qquad {}\times J_{\Omega}(x_3) (i\slashed{\partial} -\Sigma)^{-1} \delta(x_3-x_1) \bar H(x_1) \nonumber\\
&&\qquad\qquad {} \times\left(i v\cdot\partial-\frac{(\slashed \partial_{\perp})^2 }{2m_Q}\right)^{-1} \delta(x_1-x_2)  H (x_2)\Bigg].\nonumber\\
\label{S3}
\end{eqnarray}

The external source $J_{\Omega}$ contains the scalar part $S_\Omega$ and the pseudo-scalar part $P_\Omega$, both of which are of $\mathcal{O}(p^2)$ in the chiral power counting. We only consider up to $\mathcal{O}(p)$ order chiral Lagrangian, so that their contributions are ignored. Then we obtain
\begin{widetext}
\begin{eqnarray}
S_3 & = & S^{(0)}_{3}+S^{(1)}_{3},\nonumber\\
S^{(0)}_{3} & = & {} -iN_c\int d^4x \int\frac{d^4 p}{(2\pi)^4}\left[\frac{\Sigma^2+\frac{1}{3}p^2}{(p^2-\Sigma^2) v\cdot p}  - \frac{2\Sigma}{(p^2-\Sigma^2)^2 } \right] \mathrm{Tr}\left[ H(x) \gamma_{\mu}\gamma_5 A^{\mu}_{\Omega}(x) \bar H(x)\right]\nonumber\\
&&{} -iN_c\int d^4x \int\frac{d^4 p}{(2\pi)^4} \left[\frac{1}{(p^2-\Sigma^2) v\cdot p} +\frac{2\Sigma }{(p^2-\Sigma^2)^2} \right]  \mathrm{Tr}\left[ H(x) v_{\mu} V^{\mu}_{\Omega}(x) \bar H(x)\right],\nonumber\\
S^{(1)}_{3} & = &{} -iN_c\int d^4x \int\frac{d^4 p}{(2\pi)^4} \left[\frac{\Sigma^2+\frac{1}{3}p^2}{(p^2-\Sigma^2)^2 v\cdot p}  - \frac{2\Sigma-\frac{2}{3}v\cdot p}{(p^2-\Sigma^2)^2 } \right]\biggr[\frac{-p^2+ (v\cdot p)^2}{2 m_Q v\cdot p }\biggr]\mathrm{Tr}\left[ H(x) \gamma_{\mu}\gamma_5 A^{\mu}_{\Omega}(x) \bar H(x)\right]  \nonumber\\
&&{} -iN_c\int d^4x \int\frac{d^4 p}{(2\pi)^4} \left[\frac{1}{(p^2-\Sigma^2) v\cdot p} +\frac{2\Sigma-2v\cdot p }{(p^2-\Sigma^2)^2} \right] \biggr[\frac{-p^2+ (v\cdot p)^2}{2 m_Q v\cdot p }\biggr]  \mathrm{Tr}\left[ H(x) v_{\mu} V^{\mu}_{\Omega}(x) \bar H(x)\right].
\label{eq:S33}
\end{eqnarray}
Again, $S^{(0)}_{3}$ represents the leading order term in the $1/m_Q$ expansion and $S^{(1)}_{3}$ is the $\mathcal{O}(1/m_Q)$ term.

Summing up Eqs.~\eqref{eq:S22} and \eqref{eq:S33}, we obtain the expressions for the constants $m_H$ and $g_H$  as
\begin{eqnarray}
  m_H & = & \frac{iN_c}{Z_H}  \int\frac{d^4 p}{(2\pi)^4}   \left[{}\frac{1}{p^2-\Sigma^2} - \frac{\Sigma}{(p^2-\Sigma^2)v\cdot p}\right] \biggr[1+\frac{-p^2+ (v\cdot p)^2}{2 m_Q v\cdot p }\biggr]  ,  \nonumber\\
  g_{H} & = &{} -\frac{iN_c}{Z_H}\int\frac{d^4 p}{(2\pi)^4} \left[\frac{\Sigma^2+\frac{1}{3}p^2}{(p^2-\Sigma^2)^2 v\cdot p}  - \frac{2\Sigma}{(p^2-\Sigma^2)^2 } \right]\biggr[1+\frac{-p^2+ (v\cdot p)^2}{2 m_Q v\cdot p }\biggr] +\frac{\frac{2}{3}[-p^2+ (v\cdot p)^2] } {2m_Q (p^2-\Sigma^2)^2}.
  \label{eq:mHgH}
\end{eqnarray}
One can find that the coefficients of $\mathrm{Tr}\left[ H(x)iv\cdot\partial \bar H(x)\right]$ term and $\mathrm{Tr}\left[ H(x)v\cdot V_\Omega \bar H(x)\right]$ term are the same. This means that the vector part of the chiral symmetry is reserved in our approach, in agreement with the pattern of the chiral symmetry breaking in QCD. According to our previous definition of the covariant derivative $\nabla$, we can write them uniformly as $\mathrm{Tr}\left[ H(x)iv\cdot\nabla \bar H(x)\right]$. The coefficient of this term is called the wave function renormalization factor which is given by
\begin{eqnarray}
Z_H & = & iN_c\int\frac{d^4 p}{(2\pi)^4}\biggr[{} -\frac{1}{(p^2-\Sigma^2) v\cdot p} - \frac{2\Sigma+2\Sigma'( p^2+\Sigma^2)}{(p^2-\Sigma^2)^2}\biggr] \biggr[1+\frac{-p^2+ (v\cdot p)^2}{2 m_Q v\cdot p }\biggr] + \frac{(4\Sigma\Sigma'+2)[-p^2+ (v\cdot p)^2]}{2m_Q (p^2-\Sigma^2)^2}.
\end{eqnarray}

Following the same procedure, we obtain the LECs for the field $G$ and the coupling constant between the chiral partner fields $H$ and $G$, $g_{HG}$, as
\begin{eqnarray}
	m_G & = & \frac{iN_c}{Z_G}\int\frac{d^4 p}{(2\pi)^4} \left[\frac{1}{p^2-\Sigma^2} +\frac{\Sigma}{(p^2-\Sigma^2)v\cdot p}\right] \biggr[1+\frac{-p^2+ (v\cdot p)^2}{2 m_Q v\cdot p }\biggr] , \nonumber\\
	g_{G} & = &{} -\frac{iN_c}{Z_G}\int\frac{d^4 p}{(2\pi)^4}  \left[\frac{\Sigma^2+\frac{1}{3}p^2}{(p^2-\Sigma^2)^2 v\cdot p}  + \frac{2\Sigma}{(p^2-\Sigma^2)^2 } \right] \biggr[1+\frac{-p^2+ (v\cdot p)^2}{2 m_Q v\cdot p }\biggr] +\frac{\frac{2}{3}[-p^2+ (v\cdot p)^2] } {2m_Q (p^2-\Sigma^2)^2} \nonumber\\	
	Z_G & = & iN_c\int\frac{d^4 p}{(2\pi)^4}\biggr[{} -\frac{1}{(p^2-\Sigma^2) v\cdot p} +\frac{2\Sigma+2\Sigma'( p^2+\Sigma^2)}{(p^2-\Sigma^2)^2} \biggr]\biggr[1+\frac{-p^2+ (v\cdot p)^2}{2 m_Q v\cdot p }\biggr]+ \frac{(4\Sigma\Sigma'+2)[-p^2+ (v\cdot p)^2]}{2m_Q (p^2-\Sigma^2)^2}.\nonumber\\
g_{H G} & = & {} - i \frac{N_c}{\sqrt{Z_H Z_G}}\int\frac{d^4 p}{(2\pi)^4}  \biggl[\frac{\Sigma^2+p^2}{(p^2-\Sigma^2)^2 v\cdot p} \biggr] \biggr[1+\frac{-p^2+ (v\cdot p)^2}{2 m_Q v\cdot p }\biggr] -\frac{2 [-p^2+ (v\cdot p)^2] } {2m_Q (p^2-\Sigma^2)^2}  .\label{eq:gHG}
\end{eqnarray}
\end{widetext}

So far, we have obtained all the LECs of heavy-light mesons chiral effective Lagrangian with $1/m_Q$ correction in Eqs.~\eqref{eq:mHgH} - \eqref{eq:gHG}. The LECs are expressed as integrals of the light-quark self-energy $\Sigma(-p^2)$ and its derivative, which can be calculated from QCD.

In deriving the effective action for the chiral effective Lagrangian, we have taken a dynamical expansion which retains only the minimum of QCD dynamics responsible for generating the dynamical chiral symmetry breaking. That's why the Gluon effects are taken into account only through the quark dressing functions. This expansion has been found
workable well for the calculation of the pion decay constant~\cite{Pagels:1979hd}. This simplification unfortunately has a consequence that the mass splitting between the spin-0 and spin-1 particles cannot be generated --- the magnetic momentum operator of the quark-gluon interaction cannot be generated. However, corrections could be systematically calculated by taking into account higher order gluon Green functions, which is beyond our present purpose.

\section{Numerical results }
\label{sec:Num}
Given the expressions of LECs shown in the previous section, one can calculate the masses of the heavy-light mesons and the coupling constants, as long as the light quark self-energy $\Sigma(-p^2)$ is properly obtained. Following our previous work~\cite{Chen2020jiq}, we use the self-energy obtained from the Dyson-Schwinger equations as well as the lattice QCD to perform the numerical calculations.

For the Dyson-Schwinger equation method, we use the differential form of the gap equation~\cite{Yang:2002hea}:
\begin{eqnarray}\label{gap}
& &\left(\frac{\alpha_s(x)}{x}\right)'\Sigma(x)''-\left(\frac{\alpha_s(x)}{x}\right)'' \Sigma(x)' \nonumber\\
&&\qquad\qquad\qquad\quad {}- \frac{3C_2(R)}{4\pi}\frac{x\Sigma(x)}{x+\Sigma^2(x)}
\left(\frac{\alpha_s(x)}{x}\right)'^2=0,\nonumber
\end{eqnarray}
with boundary conditions
\begin{eqnarray}
\Sigma'(0) & = &{} - \frac{3C_2(R)\alpha_s(0)}{8\pi\Sigma(0)}, \nonumber\\
\Sigma(\Lambda') & = & \frac{3C_2(R)\alpha_s(\Lambda')}{4\pi\Lambda'}
\int^{\Lambda'}_0dx\frac{x\Sigma(x)}{x+\Sigma^2(x)},
\end{eqnarray}
where $\alpha_s$ is the running coupling constant of QCD. $\Lambda'$ is an ultraviolet cutoff regularizing the integral, which should be taken $\Lambda^\prime \rightarrow \infty$ eventually. Since the low energy behavior of the strong coupling constant $\alpha_s$ is not clear up to now, we take a model description for $\alpha_s$ given in Ref.~\cite{Dudal:2012zx}, which is called the refined Gribov-Zwanziger (G-Z) formalism:~\footnote{In our previous work~\cite{Chen2020jiq}, another model was also adopted. However the model gives similar results as those by using the self-energy function fitted from Lattice QCD. So in this work, we shall not consider that model.}
\begin{equation}
\alpha_s(p^2)=\alpha_0p^2\frac{p^2+M^2}{p^4+(M^2+m^2)p^2+\lambda^4},
\end{equation}
where $M^2=4.303~{\rm GeV}^2 , (M^2+m^2)= 0.526~{\rm GeV}^2 $ and $\lambda^4=0.4929~{\rm GeV}^4$. $\alpha_0$ is a model parameter to be determined. Although this model does not respect the UV behavior of QCD, since the LECs are mostly controlled by the low energy behavior of QCD, this should not be a problem in our study.

\begin{figure}[h]
	\centering	
	\includegraphics[height=5.0cm]{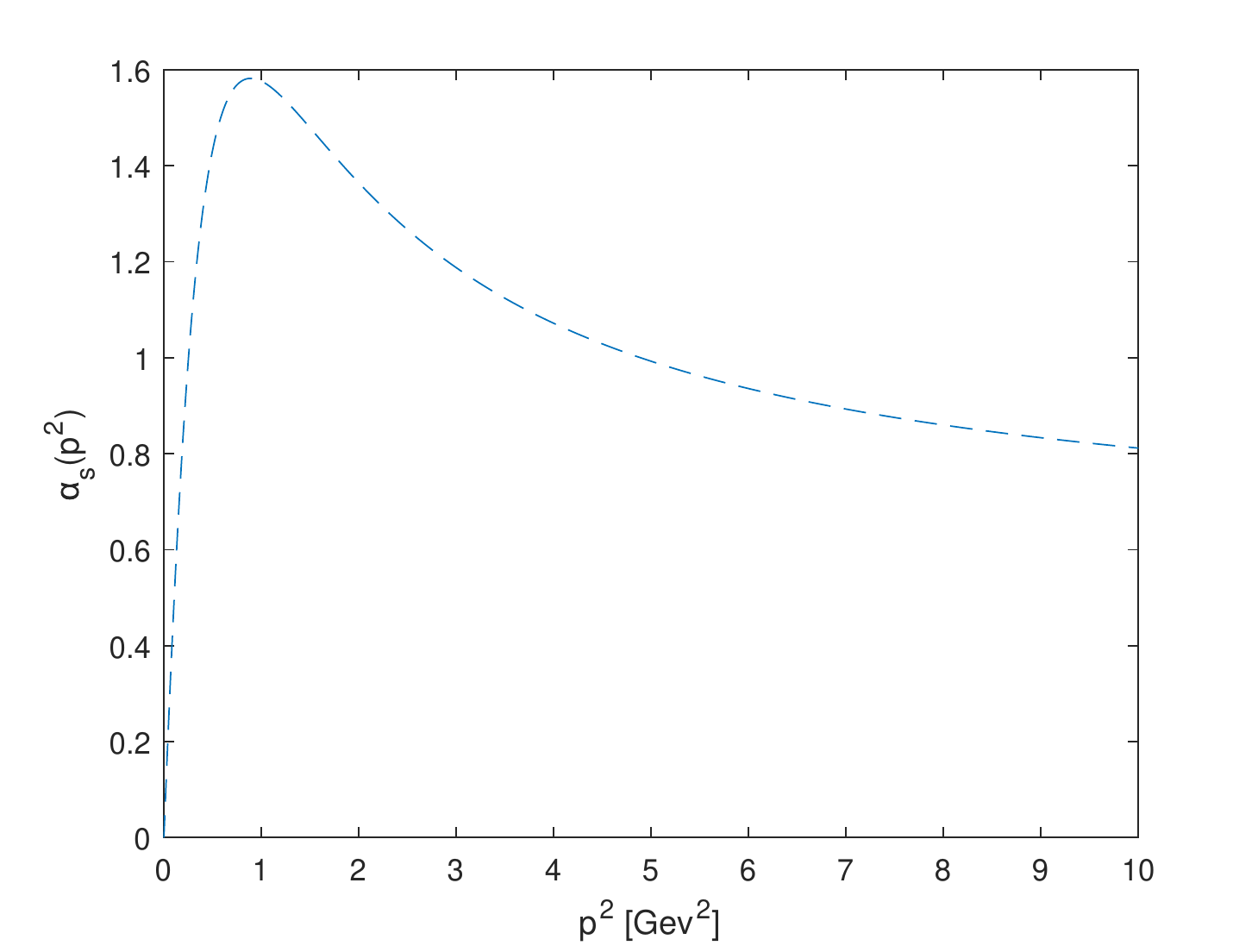}
	\caption{Running coupling constant of the G-Z model with $a_0=0.60$. }	
	\label{rcc}
\end{figure}

\begin{figure}[h]
	\centering	
	\includegraphics[height=5cm]{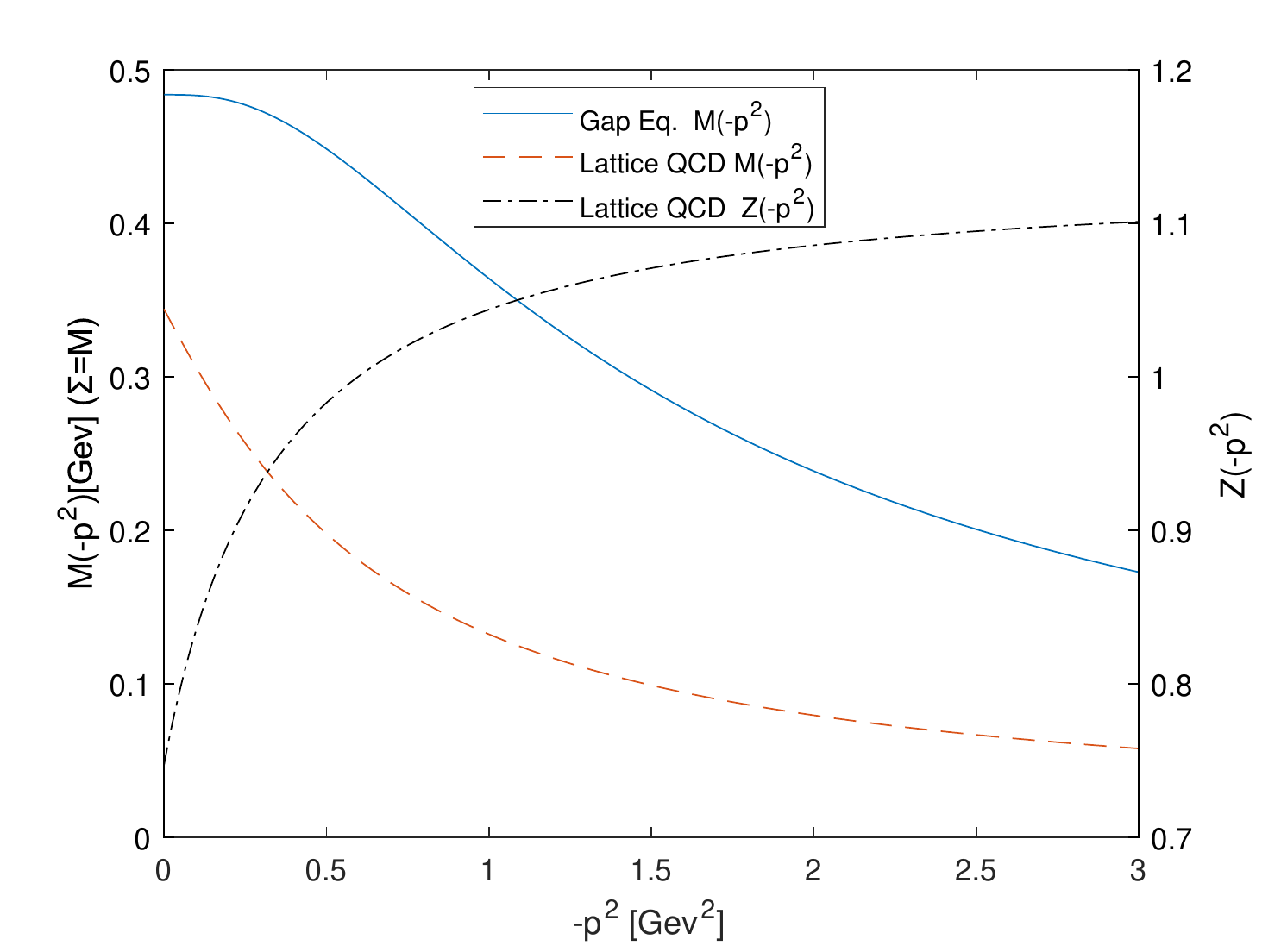}
	\caption{The lattice fittings of $M(-p^2)$ and $Z(-p^2)$ given in Ref. \cite{Oliveira:2018ukh} and $\Sigma(-p^2)$ from the gap equation with Model G-Z. }	
	\label{selfenergy}
\end{figure}

The LECs are calculated according to Eqs.~\eqref{eq:mHgH}-\eqref{eq:gHG}. It is clear that the integrals of the LECs have a physical ultraviolet cutoff $\Lambda_c$ which should be of the order of the chiral symmetry breaking scale and serves as another parameter in our calculations. When studying the low energy constants in the heavy quark limit, we have found that there is a tension between the pion decay constant $f_\pi$ and the mass splitting of the chiral partners $\Delta m$, i.e., no parameter sets could fit both quantities perfectly~\cite{Chen2020jiq}. In this work, we find that the tension is largely released as long as the $1/m_Q$ corrections are taken into account. Since $f_\pi$ is a well-established quantity, we shall take $f_\pi\simeq 93$~MeV as an input to determine $\Lambda_c$, and leave the parameter $a_0$ as a free one. The parameter $a_0$ is scanned from $a_0=0.52$ through $a_0=0.62$, and we find that $a_0=0.60$ gives the best fitted results. To give an intuitive impression, we draw the running coupling constant $\alpha_s$ calculated with the G-Z formalism at $a_0=0.60$ in Fig.~\ref{rcc}. The light quark self-energy $\Sigma(-p^2)$ solved by the gap equation \eqref{gap} is shown in Fig.~\ref{selfenergy}. In Table~\ref{tab1}, we list the LECs with and without $1/m_{Q}$ corrections.

\begin{table*}[!htbp]
	\centering
	\caption{The heavy-light meson masses and the coupling constants calculated from the G-Z Model with $a_0=0.6$. }
	\begin{tabular}{cccccccccc}
		\hline	\hline
		Correction &  $\Sigma(0)$ (GeV) &  $f_{\pi}$ (GeV) & ${}-\langle\bar{\psi}\psi\rangle$ (GeV$^{3}$)& $\quad$ $g_G$ $\quad$  & $\quad$ $g_H$ $\quad$ & $\quad$ $g_{HG}$ $\quad$ & $m_G$~(GeV) & $m_H$~(GeV) & $\Delta m $~(GeV)\\
		\hline
	      LO    & 0.484 & 0.093 & 0.261 & 1.570 & -0.576 & 0.022 & 0.963 & 0.662 & 0.301 \\
        $1/m_c$  & 0.484 & 0.093 & 0.261 & 1.862 & -0.554 & -0.139 & 1.142 & 0.739 & 0.403 \\
        $1/m_b$  &	0.484 & 0.093 & 0.261 & 1.649 & -0.570 & -0.022 & 1.012 & 0.683 & 0.329 \\
		\hline	\hline
	\end{tabular}%
	\label{tab1}%
\end{table*}%

From Table~\ref{tab1}, we find that the $1/m_Q$ corrections for the LECs except $g_{HG}$ are ranging from $\sim 1\% - 5\%$ for bottom mesons and from $\sim 4\% - 19\%$ for charmed mesons. This agrees with the fact that, compare to the charm quark sector, the heavy quark expansion is more valid in the bottom quark sector. It is interesting to notice that the corrections to the LECs associated to $H$ fields are typically smaller than those associated to $G$ fields by a few percents. It is found that the mass splitting of the chiral partners $\Delta m$ which is explicitly related to the dynamical chiral symmetry breaking of QCD in our method suffers a significant correction for the charmed mesons, and the resulted value is consistent with the experimental data (recalling that the averaged mass splitting for the $D$ meson sector is $\Delta m=0.430$ GeV.) The mass splitting for the $B$ meson sector suffers the $1/m_b$ correction by about $9\%$. According to our results, we may conclude that, for the charmed mesons, the $1/m_Q$ corrections usually contribute significantly, while for the bottom mesons, taking the heavy quark limit may cause errors up to $\sim 10$ percents.

The masses $m_H$ and $m_G$ displayed in Table~\ref{tab1} are the ``residue masses" with the heavy quark mass rotated away. The physical masses for the $H$ and $G$ doublets can be easily obtained by restoring the heavy quark mass $m_c$ or $m_b$. Using $m_c\approx 1.27$ GeV and $m_b\approx 4.66$ GeV~\cite{ZylaPDG}, we obtain the physical masses for the $D$ meson family to be $\tilde{m}_H\approx 2.01$ GeV and $\tilde{m}_G\approx 2.41$ GeV; and for the $B$ meson family to be $\tilde{m}_H\approx 5.34$ GeV and $\tilde{m}_G\approx 5.67$ GeV. These results are consistent with the experimental data for the charmed mesons~\eqref{eq:EXmass} and the bottom mesons~\eqref{eq:EXmass'} respectively.

The coupling constant $g_H$ directly governs the decay process $D^{\ast} \to D \pi $. Given the value shown in Table~\ref{tab1}, we find the decay width
\begin{eqnarray}
\Gamma(D^{\ast +}\to D\pi)=\frac{3}{2}\frac{g^2_{H} |P_{\pi}|^3}{12\pi f_\pi^2}=87.2~\text{KeV},
\end{eqnarray}
which is very close to the experimental result $ 83.4 \pm 1.8$~KeV~\cite{ZylaPDG}.


We now turn to the calculation using the light quark self-energy obtained from lattice QCD. In Ref.~\cite{Oliveira:2018ukh}, the authors fitted the lattice results for the quark wave function renormalization $Z(-p^2)$ and the running quark mass $M(-p^2)$ (corresponding to $\Sigma(-p^2)$ in our terminology). For illustration, we plot $M(-p^2)$ and $Z(-p^2)$ in Fig.~\ref{selfenergy}. In the previous section, we present the formula for LECs with the wave function renormalization $Z(-p^2)$ ignored. It is straightforward to keep $Z(-p^2)$ in the formula which are given in Appendix B. For comparison, we performed the calculations using the fitted functions from lattice QCD with $Z(-p^2)$ (see Table~\ref{tab2}) and without $Z(-p^2)$ (see Table~\ref{tab3}). One can see from Table~\ref{tab3} that the mass splitting is significantly deviated from the expected values in the $Z(-p^2)=1$ case. The reason may be understood that in the lattice QCD, both $M(-p^2)$ and $Z(-p^2)$ are needed to describe the dressing effects of the quark propagator, while in the G-Z model, the running quark mass $M(-p^2)$ (i.e., $\Sigma(-p^2)$ in the G-Z model) alone satisfies the Dyson-Schwinger equation for the quark propagator.
\begin{table*}[!htbp]
	\centering
	\caption{LECs calculated from lattice fittings given in Ref \cite{Oliveira:2018ukh}. }
	\begin{tabular}{cccccccccc}
		\hline	\hline
		Correction & $M(0)$(GeV) 	& $f_{\pi}$ (GeV) & ${}-\langle\bar{\psi}\psi\rangle$ (GeV$^{3}$) & $\quad$ $g_G$ $\quad$  & $\quad$ $g_H$ $\quad$ & $\quad$ $g_{HG}$ $\quad$ & $m_G$~(GeV) & $m_H$~(GeV) & $\Delta m $~(GeV)\\
		\hline
		LO    & 0.418 & 0.093 & 0.307 & 0.820 & -0.095 & 0.805 & 1.012 & 0.660 & 0.352 \\
		$1/m_c$ & 0.418 & 0.093 & 0.307 & 1.116 & -0.032 & 0.696 & 1.246 & 0.806 & 0.440 \\
		$1/m_b$ & 0.418 & 0.093 & 0.307 & 0.901 & -0.078 & 0.776 & 1.076 & 0.700 & 0.376 \\
		\hline	\hline
	\end{tabular}%
	\label{tab2}%
\end{table*}%

\begin{table*}[!htbp]
	\centering
	\caption{LECs calculated from lattice fittings given in Ref \cite{Oliveira:2018ukh} with $Z(-p^2)=1$. }
	\begin{tabular}{cccccccccc}
		\hline	\hline
		Correction & $M(0)$(GeV) 	& $f_{\pi}$ (GeV) & ${}-\langle\bar{\psi}\psi\rangle$ (GeV$^{3}$) & $\quad$ $g_G$ $\quad$  & $\quad$ $g_H$ $\quad$ & $\quad$ $g_{HG}$ $\quad$ & $m_G$~(GeV) & $m_H$~(GeV) & $\Delta m $~(GeV)\\
		\hline
		  LO    & 0.418 & 0.093 & 0.307 & 1.016 & -0.137 & 0.819 & 1.183 & 0.611 & 0.572 \\
		 $1/m_c$ & 0.418 & 0.093 & 0.307 & 1.556 & -0.087 & 0.749 & 1.657 & 0.710 & 0.947 \\
		$1/m_b$  & 0.418 & 0.093 & 0.307 & 1.163 & -0.123 & 0.800 & 1.312 & 0.638 & 0.674 \\
		\hline	\hline
	\end{tabular}%
	\label{tab3}%
\end{table*}%

From Table~\ref{tab2}, we see that the $1/m_Q$ corrections to the LECs are even larger compared to the G-Z model results. The mass splitting for $D$ and $B$ mesons is consistent with the experimental results respectively. The physical masses are $\tilde{m}_H\approx 2.08 $ GeV and $\tilde{m}_G\approx 2.52$ GeV for $D$ mesons, and $\tilde{m}_H\approx 5.36$ GeV and $\tilde{m}_G\approx 5.74$ GeV for $B$ mesons. The masses of $D$ mesons are slightly larger than the experimental data (Eq.~\eqref{eq:EXmass}), while the masses of $B$ mesons are consistent with the  experimental data (Eq.~\eqref{eq:EXmass'}). We also notice from Table~\ref{tab2} that the coupling constant $g_H$ is too small to fit the experimental data, which indicates the shortcomings of the results based on lattice fittings. In general, coupling constants in the effective Lagrangian are more sensitive to the details of $M(-p^2)$ and $Z(-p^2)$ than the masses are. Since the functions $Z(-p^2)$ and $M(-p^2)$ are fitted from the lattice data corresponding to $M_\pi=295$ MeV~\cite{Oliveira:2018ukh}, we are not expecting them could reproduce every LEC properly.

\section{Summary}
\label{sec:dis}

In this paper, we extend our previous work on deriving the chiral effective Lagrangian for heavy-light mesons from QCD to include $1/m_{Q}$ corrections. Using the quark self-energy and the wavefunction renormalization calculated from Dyson-Schwinger equations as well as those fitted from lattice QCD, we calculated the $1/m_Q$ corrections to the LECs in the Lagrangian. It is found that for charmed mesons, the $1/m_Q$ corrections are significant and typically give about $\sim 20\%$ contributions, while for bottom mesons, the $1/m_Q$ corrections are usually ranging from a few percents up to $\sim10\%$.

The $1/m_{Q}$ order corrections improve the leading order results. At the leading order in the $1/m_Q$ expansion, the pion decay constant $f_\pi$ and the mass splitting of the chiral partners $\Delta m$ cannot be fitted to the corresponding experimental data simultaneously. However, with the $1/m_Q$ contributions included, our numerical results from both the G-Z model and the lattice fittings are comparable to the corresponding experimental data. In addition, in the G-Z model, most of the LECs at $1/m_Q$ order are consistent with the existent experimental results. Moreover, it is also find that the $D^\ast \to D \pi$ decay width is improved to $\Gamma(D^{\ast +}\to D\pi)=87.2~\text{KeV}$ which is consistent to the experimental result.

\acknowledgments
The work of Y.~L. M. was supported in part by National Science Foundation of China (NSFC) under Grant No. 11875147 and No.11475071. Q. W. was supported by the National Science Foundation of China (NSFC) under Grant No. 11475092.

\appendix

\renewcommand{\appendixname}{Appendix}

\section{DERIVATION OF THE ACTION}\label{sec:AppA}
We derive the effective action at the $1/m_{Q}$ order from QCD in this appendix.

\subsection{Integrate in the Nambu-Goldstone boson fields  }
Starting from Eq.~(\ref{generating2}) in Sec.~\ref{sec:EFTQCD}, we now introduce the pseudoscalar meson field $U$ into the theory. Inserting the following constant into the generating functional
\begin{eqnarray}\label{con1}
I & = &\int \mathcal{D}U \delta(U^{\dagger}U-1)\delta(\det U-1)\nonumber\\
& & \quad{} \times \mathcal{F}[\mathcal{O}]\delta\left(\Omega\mathcal{O}^{\dagger}\Omega-\Omega^{\dagger}\mathcal{O}\Omega^{\dagger}\right),\nonumber\\
&&
\end{eqnarray}
where  $\mathcal{F}[\mathcal{O}]\equiv\{\prod\limits_x\det\mathcal{O}\int\mathcal{D}\sigma\delta(\mathcal{O}^{\dagger}\mathcal{O}-\sigma^{\dagger}\sigma)\delta(\sigma-\sigma^{\dagger})\}^{-1}$, $\mathcal{O}(x)=e^{-i[\theta(x)/N_f]}\mathrm{tr}_l[P_R B^T(x,x)]$, and $B$ is the abbreviation of the bilocal composite light quark fields. Then, integrating out the gluon fields, we obtain
\begin{widetext}
\begin{eqnarray}\label{generating4}
Z[J] & = & \int \mathcal{D} q \mathcal{D} \bar q \mathcal{D} N_v \mathcal{D}\bar N_v \mathcal{D}\bar U \delta(\mathcal{O}^{\dagger}-\mathcal{O}) \nonumber\\
& &{} \times \exp\biggl\{i\int d^4x \biggl[ \bar  q (i\slashed{\partial}+J_\Omega) q + \bar N_v \left ( i v\cdot \partial -\frac{\partial^2-(v\cdot\partial)^2 }{2m_Q} \right ) N_v \biggr]+i\Gamma_I[B] \notag \\
& &\qquad\quad{} +\sum\limits_{n=2}^{\infty}\int d^4x_1 d^4x'_1 \cdots d^4x_n d^4 x'_n \nonumber\\
& &\qquad\qquad\qquad{}\times \dfrac{(-i)^n (N_cg^2)^{n-1}}{n!} \bar G^{\sigma_1 \cdots \sigma_n}_{\rho_1 \cdots \rho_n}(x_1,x'_1, \cdots ,x_n,x'_n) \bar {\psi}^{\sigma_1}_{\alpha_1}(x_1)   \psi^{\rho_1}_{\alpha_1}(x'_1)\cdots \bar {\psi}^{\sigma_n}_{\alpha_n}(x_n) \psi^{\rho_n}_{\alpha_n}(x'_n) \biggl\},\notag
\end{eqnarray}
\end{widetext}
where the Fierz reordering has been made and the extended gluon Green functions $\bar G^{\sigma_1 \cdots \sigma_n}_{\rho_1 \cdots \rho_n}(x_1,x'_1, \cdots ,x_n,x'_n)$ have been introduced, $\mathcal{D}\bar U =\mathcal{D}U \delta(U^{\dagger}U-1)\delta(\det U-1) $ and $e^{i\Gamma_I[B]}\equiv \mathcal{F}[\mathcal{O}]$. The details of Fierz reordering and introducing the extended gluon Green functions could be found in Refs~\cite{Wang:1999cp,Fu:2017azw}. Due to the $1/m_Q$ corrections, $\bar{G}$'s include more rich types of gluon Green functions than regular ones.
The field $\psi$ stands for both the light and heavy quarks, i.e., $\psi=(q,N_v)$. We have made chiral rotation of the original light quark fields and ignored the chiral anomaly. $J_{\Omega}(x)$ is the external source under chiral rotation~\cite{Wang:1999cp}
\begin{eqnarray}
J_{\Omega}(x)&=&[\Omega(x)P_{R}+\Omega^{\dagger}(x)P_L] \notag\\
&&\times [J(x)+i\slashed \partial][\Omega(x)P_{R}+\Omega^{\dagger}(x)P_L].
\end{eqnarray}

\subsection{Integrate in the heavy-light meson fields}
The heavy-light meson fields may be introduced by inserting the following constant
\begin{equation}
\int \mathcal{D}\Phi\delta(N_c\Phi^{ab}_{\eta\xi}(x,x')-\bar{ \psi}^{a}_{\eta}(x)  \psi^{b}_{\xi}(x')),
\end{equation}
into Eq.~\eqref{generating4}, where $\Phi^{ab}_{\eta \xi }$ is a bilocal auxiliary field.
The $\delta$ function  can be further expressed in the
Fourier representation
\begin{widetext}
\begin{eqnarray}
	\delta{(N_c\Phi^{ab}_{\eta\xi}(x,x')-\bar{\psi}^{a}_{\eta}(x) \psi^{b}_{\xi}(x') )}\sim \int \mathcal{D}\Pi \exp^{i\int d^4 xd^4 x'\Pi^{ab}_{\eta\xi}(x,x')[N_c\Phi^{ab}_{\eta\xi}(x,x')-\bar{\psi}^{a}_{\eta}(x) \psi^{b}_{\xi}(x')]}.
	\label{delta}
\end{eqnarray}
Then the generating functional becomes
\begin{eqnarray}
	Z[J] & = & \int \mathcal{D} q\mathcal{D} \bar q\mathcal{D}  N_v \mathcal{D}\bar{N}_v\mathcal{D}\bar U\mathcal{D}\Phi\mathcal{D}\Pi  \delta(\mathcal{O}^{\dagger}-\mathcal{O}) \notag\\
	& & {} \times\exp\biggl\{i\int d^4x \biggl[ \bar  q (i\slashed{\partial}+J_\Omega) q+ \bar N_v \left ( i v\cdot \partial -\frac{\partial^2-(v\cdot\partial)^2 }{2m_Q} \right ) N_v\biggr] +i\Gamma_I[\Phi]+iN_c\bar G(\Phi)\notag\\
	& & {} \qquad\qquad  +i\int d^4 xd^4 x'\biggl[N_c\Pi^{ab}_{\eta\xi}(x,x')\Phi^{ab}_{\eta\xi}(x,x')-\bar{\psi}(x)\Pi(x,x') \psi(x')\biggr] \biggr\},
	\label{generating6}
\end{eqnarray}
where
\begin{eqnarray}
	\bar G(\Phi)=\sum\limits_{n=2}^{\infty}\int d^4x_1 d^4x'_1 \cdots d^4x_n  d^4 x'_n \dfrac{(-i)^n (N_cg^2)^{n-1}}{n!} \bar G^{\sigma_1 \cdots \sigma_n}_{\rho_1 \cdots \rho_n}(x_1,x'_1, \cdots ,x_n,x'_n) \Phi^{\sigma_1\rho_1}(x_1,x'_1) \cdots \Phi^{\sigma_n\rho_n}(x_n,x_n').\nonumber
\end{eqnarray}

By integrating out the quark fields $\psi$ and $\bar{\psi}$, we obtain
\begin{eqnarray}
	Z[J] & = & \int\mathcal{D}\bar U\mathcal{D}\Phi\mathcal{D}\Pi  \delta(\mathcal{O}^{\dagger}-\mathcal{O}) \nonumber\\
	& &{} \times\exp\biggl\{iN_c\biggl[-i \mathrm{Tr}^\prime \ln[ i\slashed{\partial} I_1+J'_\Omega +  \left ( i v\cdot \partial -\frac{\partial^2-(v\cdot\partial)^2 }{2m_Q} \right ) I_4 -\Pi ]\biggr] +iN_c \mathrm{Tr}^\prime(\Pi\Phi^T) +iN_c\bar G(\Phi) + i\Gamma_I[\Phi] \biggr\},\notag\\
	\end{eqnarray}
\end{widetext}
where we have defined the functional trace $\mathrm{Tr}^\prime$ taking over the flavor space, spinor space and coordinate space, $I_1=\mathrm{diag}(1,1,0)$ and $I_4=\mathrm{diag}(0,0,1)$ are the  matrices in the flavor space.  $J'_\Omega$ is the extension of $J_\Omega$ to the whole flavor space.

In flavor space the matrix $\Pi$ can be decomposed as
\begin{eqnarray}
\Pi=\Pi_1+\Pi_2+\Pi_3+\Pi_4,
\end{eqnarray}
where
\begin{eqnarray}
\Pi_1=\begin{pmatrix}
\Pi^{qq}_{2\times2}  & 0\\
0 & 0
\end{pmatrix}, \quad \Pi_2=\begin{pmatrix}
0  & 0\\
\Pi^{Qq}_{1\times2} & 0
\end{pmatrix},
\nonumber\\
\quad \Pi_3=\begin{pmatrix}
0  & \Pi^{qQ}_{2\times1}\\
0 & 0
\end{pmatrix},\quad \Pi_4=\begin{pmatrix}
0 & 0\\
0 & \Pi^{QQ}
\end{pmatrix}.
\end{eqnarray}
Given this decomposition, we recognize that $\Pi_2$ and $\Pi_3$ are the bosonic interpolating fields for the heavy-light mesons. To get a local effective Lagrangian, we take the following localization conditions which are essentially the point coupling between quarks~\cite{Tandy:1997qf}
\begin{eqnarray}
\Pi_2(x,y) & = & \Pi_2(x)\delta(x-y), \nonumber\\
\Pi_3(x,y) & = & \Pi_3(x)\delta(x-y).
\end{eqnarray}
It is easy to check that $\bar{\Pi}_2(x,x)\equiv\gamma^0\Pi_2^{\dagger }(x,x)\gamma^0=\Pi_{3}(x,x)$, then the generating functional can be written as
\begin{widetext}
	\begin{eqnarray}
	\label{gf}
	Z[J] & = & \int\mathcal{D}\bar U\mathcal{D}\Phi\mathcal{D}\Pi  \exp\Biggl\{i\Gamma_I[\Phi]+iN_c\biggr\{\mathrm{Tr}^\prime(\Pi\Phi^T) +\bar G(\Phi)+\int d^4x\mathrm{Tr}\left\{\Xi\left[\left(-i\sin\frac{\theta}{N_f}+\cos\frac{\theta}{N_f}\right)\Phi^T\right]\right\}\notag\\
	& &{}\qquad\qquad\qquad\qquad\qquad -i \mathrm{Tr}^\prime \ln\left[i\slashed{\partial}I_1 +J'_\Omega
	+ \left ( i v\cdot \partial -\frac{\partial^2-(v\cdot\partial)^2 }{2m_Q} \right ) I_4  -\Pi_1-\Pi_2-\bar{\Pi}_2 -\Pi_4\right]\biggr\} \Biggr\}
	\end{eqnarray}
\end{widetext}
where the $\delta(\mathcal{O}^{\dagger}-\mathcal{O}) $ term has been reexpressed as
\begin{eqnarray}
\delta(\mathcal{O}^{\dagger}-\mathcal{O})=\int \mathcal{D}\Xi e^{iN_c\int d^4 x \mathrm{Tr}\{\Xi (x)[\Theta\Phi^T(x,x)]\}}
\end{eqnarray}
with $\Xi^{ab}(x)$ being a new auxiliary field and $\Theta  \equiv (-i\sin\theta(x)/N_f+\cos\theta(x)/N_f).$

Now, by integrating out the fields $\Phi$, $\Xi$, $\Pi_1$ and $\Pi_4$, we can obtain the action, denoted as $S[U,\Pi_2,\bar{\Pi}_2]$, for the chiral effective theory with heavy-light mesons
\begin{eqnarray}
Z[J]=\int\mathcal{D}\bar U\mathcal{D}\Pi_2\mathcal{D}\bar{\Pi}_2 \exp\{ i S[U,\Pi_2,\bar{\Pi}_2] \},
\end{eqnarray}
where
\begin{widetext}
	\begin{eqnarray}
	e^{iS} & \equiv & \int\mathcal{D}\Phi\mathcal{D}\Pi_1\mathcal{D}\Pi_4\mathcal{D}\Xi \exp\Biggl\{i\Gamma_I[\Phi]+iN_c\biggr\{\mathrm{Tr}^\prime(\Pi\Phi^T) +\bar G(\Phi)+\int d^4x\mathrm{Tr}\{\Xi [(-i\sin\frac{\theta}{N_f}+\cos\frac{\theta}{N_f})\Phi^T]\}\notag\\
	& &{}\qquad\qquad\qquad\qquad\qquad\qquad\qquad\qquad\quad -i \mathrm{Tr}^\prime \ln\left[i\slashed{\partial}I_1 +J'_\Omega
	+ \left ( i v\cdot \partial -\frac{\partial^2-(v\cdot\partial)^2 }{2m_Q} \right ) I_4  -\Pi_1-\Pi_2-\bar{\Pi}_2 -\Pi_4\right]\biggr\} \Biggr\}.\notag\\
	\end{eqnarray}
\end{widetext}

\subsection{The action in the large $N_c$ Limit }
Now, we take the large $N_c$ limit in the generating functional to obtain
\begin{widetext}
	\begin{eqnarray}
	\label{action1}
	S[U,\Pi_2,\bar{\Pi}_2] & = & N_c\Bigg\{\mathrm{Tr}^\prime(\Pi_{c}\Phi^T_c) +\bar G(\Phi_c)+\int d^4x\mathrm{Tr}\left\{\Xi_c \left[\left(-i\sin\frac{\theta(\Phi_c)}{N_f}+\cos\frac{\theta(\Phi_c)}{N_f}\right)\Phi^T_c\right]\right\}\notag\\
	&& {} \qquad -i \mathrm{Tr}^\prime \ln\left[i\slashed{\partial}I_1 +J'_\Omega
	+  \left ( i v\cdot \partial -\frac{\partial^2-(v\cdot\partial)^2 }{2m_Q} \right ) I_4  -\Pi_{1c}-\Pi_2-\bar{\Pi}_2 -\Pi_{4c}\right]\Bigg\},
	\end{eqnarray}
\end{widetext}
where $\Phi_c, \Xi_c, \Pi_{1c}, \Pi_{4c}$ are classical fields satisfying the saddle point equations
\begin{equation}
\label{saddle}
\frac{\delta S}{\delta \Phi_c}=\frac{\delta S}{\delta \Xi_c}=\frac{\delta S}{\delta \Pi_{1c}}=\frac{\delta S}{\delta \Pi_{4c}}=0.
\end{equation}
and $\Gamma_I$ term has been ignored because it is of $O(1/N_c)$~\cite{Wang:1999cp}.

These saddle point equations provide important information. For instance, equations $\frac{\delta S}{\delta \Pi_{1c}}=0$ and $\frac{\delta S}{\delta \Phi_{1c}}=0$  generate the coupled equations
\begin{equation}
\label{DS1}
\Phi^{Tab}_{1c\eta\xi}(x,y)=-i\left[(i\slashed\partial+J'_\Omega-\Pi_{1c})^{-1}\right]^{ab}_{\eta\xi}(x,y),
\end{equation}
with
\begin{widetext}
	\begin{eqnarray}
	\Pi^{ab}_{1c\eta\xi}(x,y) & = &{} -\sum\limits_{n=1}^{\infty}\int d^4x_1 \cdots d^4x'_n d^4x'_1 \cdots d^4 x'_n \dfrac{(-i)^{n+1} (N_cg^2)^{n}}{n!}\notag\\
	& &{} \qquad \qquad \times
	\bar{G}_{\eta\eta_1\cdots\eta_n,\xi\xi_1\cdots\xi_n}^{aa_1\cdots a_n,bb_1\cdots b_n}(x,y,x_1,x'_1,\cdots,x_n,x'_n) \Phi^{a_1b_1}_{1c\eta_1\xi_1}(x_1,x'_1)\cdots \Phi^{a_nb_n}_{1c\eta_n\xi_n}(x_n,x'_n),\label{DS2}
	\end{eqnarray}
\end{widetext}
where the term involving $\Xi$ has been omitted because it vanishes once the external sources $J$ are eventually turned off~\cite{Wang:1999cp}. When $J_\Omega^\prime$ is turned off, the coupled equations (\ref{DS1}) and (\ref{DS2}) are nothing but the DSEs for the quark propagators with $\Pi_{1c}$ being the self-energy for light quarks. So that we rewrite $\Pi_{1c}(x,y)$ as $\bar{\Sigma}(x-y)I_1$.
Similarly, $\Pi_{4c}$ is just the self-energy for the heavy quark. Since the contribution from the heavy quark self-energy is less significant than the light ones, we simply ignore it.

Because it is difficult to solve classical fields from the saddle point equations (\ref{saddle}) without approximations, we can not get  $U, \Pi_2$ and $\bar{\Pi}_2$  dependence of the classical fields and further obtain the chiral effective Lagrangian. Thus, we follow Ref.~\cite{Yang:2002hea} and keep only the ``$\mathrm{Tr}\ln$" term in the action in the spirit of the dynamical perturbation which works well in the calculation of pion decay constant~\cite{Pagels:1979hd}. Eventually, the effective action is simplified to be
\begin{widetext}
	\begin{eqnarray}
	\label{action2APP}
	S[U,\Pi_2,\bar{\Pi}_2] & = &
	{} -i N_c \mathrm{Tr}^\prime \ln\left[(i\slashed{\partial}-\bar{\Sigma})I_1 +J'_\Omega + \left ( i v\cdot \partial -\frac{\partial^2-(v\cdot\partial)^2 }{2m_Q} \right )  I_4-\Pi_2-\bar{\Pi}_2 \right].
	\end{eqnarray}
\end{widetext}
	
\section{FORMULA FOR LECS INCLUDING $Z(-p^2)$}
\label{sec:AppB}
In order to include the quark wave function renormalization $Z(-p^2)$  into our system, we consider the general quark propagator as
\begin{eqnarray}
S(p) & = &\frac{i}{A(-p^2) \slashed  p\ -  B(-p^2) } \nonumber\\
& = & i \, Z(-p^2) \, \frac{ \slashed  p\ + M(-p^2) }{ p^2 - M^2(-p^2)},\nonumber
\end{eqnarray}
where $Z(-p^2) = 1/A(-p^2)$ stands for the quark wave function renormalization and $M(-p^2) = B(-p^2) / A(-p^2)$ is the renormalization group invariant running quark mass. After a series of calculations, we can get the LECs with $Z(-p^2)$ as follows:

\begin{widetext}
\begin{eqnarray}
m_H & = & \frac{iN_c}{Z_H}  \int\frac{d^4 p}{(2\pi)^4} \biggr\{  \biggr[{}\frac{1}{p^2-M^2} - \frac{M}{(p^2-M^2)v\cdot p}\biggr] \biggr[1+\frac{-p^2+ (v\cdot p)^2}{2 m_Q v\cdot p }\biggr]\biggr\} Z,  \nonumber\\
g_{H} & = & -\frac{iN_c}{Z_H}\int\frac{d^4 p}{(2\pi)^4} \biggr\{  \biggl[\frac{M^2+\frac{1}{3}p^2}{(p^2-M^2)^2 v\cdot p}  - \frac{2M}{(p^2-M^2)^2 } \biggr]\biggr[1+\frac{-p^2+ (v\cdot p)^2}{2 m_Q v\cdot p }\biggr]+\frac{\frac{2}{3}[-p^2+ (v\cdot p)^2] } {2m_Q(p^2-M^2)^2}\biggr\}Z^2,\nonumber\\
m_G & = & \frac{iN_c}{Z_G}\int\frac{d^4 p}{(2\pi)^4} \biggr\{  \biggr[\frac{1}{p^2-M^2} +\frac{M}{(p^2-M^2)v\cdot p}\biggr]\biggr[1+\frac{-p^2+ (v\cdot p)^2}{2 m_Q v\cdot p }\biggr]\biggr\} Z, \nonumber\\
g_{G} & = & -\frac{iN_c}{Z_G}\int\frac{d^4 p}{(2\pi)^4} \biggr\{  \biggl[\frac{M^2+\frac{1}{3}p^2}{(p^2-M^2)^2 v\cdot p}  + \frac{2M}{(p^2-M^2)^2 } \biggr]\biggr[1+\frac{-p^2+ (v\cdot p)^2}{2 m_Q v\cdot p }\biggr]+\frac{\frac{2}{3}[-p^2+ (v\cdot p)^2] } {2m_Q(p^2-M^2)^2}\biggr\}Z^2,\nonumber\\
Z_H & = & iN_c\int\frac{d^4 p}{(2\pi)^4} \biggr\{ \biggr[ \left(-\frac{1}{(p^2-M^2) v\cdot p} - \frac{2M+2 M'(p^2+M^2)}{(p^2-M^2)^2}\right ) Z-\frac{2Z' M }{p^2-M^2} \biggr] \biggr[1+\frac{-p^2+ (v\cdot p)^2}{2 m_Q v\cdot p }\biggr] \nonumber\\
&&\qquad\qquad\qquad+ \frac{Z(4M M'+2)[-p^2+ (v\cdot p)^2]}{2m_Q (p^2-M^2)^2} +\frac{2Z'[-p^2+ (v\cdot p)^2]}{2m_Q (p^2-M^2)}\biggr\}\nonumber\\
Z_G & = & iN_c\int\frac{d^4 p}{(2\pi)^4} \biggr\{ \biggr[ \left(-\frac{1}{(p^2-M^2) v\cdot p} +\frac{2M+2 M'(p^2+M^2)}{(p^2-M^2)^2}\right ) Z+\frac{2Z' M }{p^2-M^2} \biggr] \biggr[1+\frac{-p^2+ (v\cdot p)^2}{2 m_Q v\cdot p }\biggr] \nonumber\\
&&\qquad\qquad\qquad + \frac{Z(4M M'+2)[-p^2+ (v\cdot p)^2]}{2m_Q (p^2-M^2)^2} -\frac{2Z'[-p^2+ (v\cdot p)^2]}{2m_Q (p^2-M^2)}\biggr\}\nonumber\\
g_{H G} & = &  - i \frac{N_c}{\sqrt{Z_H Z_G}}\int\frac{d^4 p}{(2\pi)^4}  \biggr\{ \biggl[\frac{M^2+p^2}{(p^2-M^2)^2 v\cdot p} \biggr]\biggr[1+\frac{-p^2+ (v\cdot p)^2}{2 m_Q v\cdot p }\biggr]-\frac{2 [-p^2+ (v\cdot p)^2] } {2m_Q(p^2-M^2)^2}\biggr\}Z^2.
\end{eqnarray}
\end{widetext}

\bibliography{paper}

\begin{thebibliography}{28}%
\makeatletter
\providecommand \@ifxundefined [1]{%
 \@ifx{#1\undefined}
}%
\providecommand \@ifnum [1]{%
 \ifnum #1\expandafter \@firstoftwo
 \else \expandafter \@secondoftwo
 \fi
}%
\providecommand \@ifx [1]{%
 \ifx #1\expandafter \@firstoftwo
 \else \expandafter \@secondoftwo
 \fi
}%
\providecommand \natexlab [1]{#1}%
\providecommand \enquote  [1]{``#1''}%
\providecommand \bibnamefont  [1]{#1}%
\providecommand \bibfnamefont [1]{#1}%
\providecommand \citenamefont [1]{#1}%
\providecommand \href@noop [0]{\@secondoftwo}%
\providecommand \href [0]{\begingroup \@sanitize@url \@href}%
\providecommand \@href[1]{\@@startlink{#1}\@@href}%
\providecommand \@@href[1]{\endgroup#1\@@endlink}%
\providecommand \@sanitize@url [0]{\catcode `\\12\catcode `\$12\catcode
  `\&12\catcode `\#12\catcode `\^12\catcode `\_12\catcode `\%12\relax}%
\providecommand \@@startlink[1]{}%
\providecommand \@@endlink[0]{}%
\providecommand \url  [0]{\begingroup\@sanitize@url \@url }%
\providecommand \@url [1]{\endgroup\@href {#1}{\urlprefix }}%
\providecommand \urlprefix  [0]{URL }%
\providecommand \Eprint [0]{\href }%
\providecommand \doibase [0]{http://dx.doi.org/}%
\providecommand \selectlanguage [0]{\@gobble}%
\providecommand \bibinfo  [0]{\@secondoftwo}%
\providecommand \bibfield  [0]{\@secondoftwo}%
\providecommand \translation [1]{[#1]}%
\providecommand \BibitemOpen [0]{}%
\providecommand \bibitemStop [0]{}%
\providecommand \bibitemNoStop [0]{.\EOS\space}%
\providecommand \EOS [0]{\spacefactor3000\relax}%
\providecommand \BibitemShut  [1]{\csname bibitem#1\endcsname}%
\let\auto@bib@innerbib\@empty
\bibitem [{\citenamefont {Wang}\ \emph
  {et~al.}(2000{\natexlab{a}})\citenamefont {Wang}, \citenamefont {Kuang},
  \citenamefont {Xiao},\ and\ \citenamefont {Wang}}]{Wang:1999cp}%
  \BibitemOpen
  \bibfield  {author} {\bibinfo {author} {\bibfnamefont {Q.}~\bibnamefont
  {Wang}}, \bibinfo {author} {\bibfnamefont {Y.-P.}\ \bibnamefont {Kuang}},
  \bibinfo {author} {\bibfnamefont {M.}~\bibnamefont {Xiao}}, \ and\ \bibinfo
  {author} {\bibfnamefont {X.-L.}\ \bibnamefont {Wang}},\ }\href {\doibase
  10.1103/PhysRevD.61.054011} {\bibfield  {journal} {\bibinfo  {journal} {Phys.
  Rev. D}\ }\textbf {\bibinfo {volume} {61}},\ \bibinfo {pages} {054011}
  (\bibinfo {year} {2000}{\natexlab{a}})}\BibitemShut {NoStop}%
\bibitem [{\citenamefont {Yang}\ \emph {et~al.}(2002)\citenamefont {Yang},
  \citenamefont {Wang}, \citenamefont {Kuang},\ and\ \citenamefont
  {Lu}}]{Yang:2002hea}%
  \BibitemOpen
  \bibfield  {author} {\bibinfo {author} {\bibfnamefont {H.}~\bibnamefont
  {Yang}}, \bibinfo {author} {\bibfnamefont {Q.}~\bibnamefont {Wang}}, \bibinfo
  {author} {\bibfnamefont {Y.-P.}\ \bibnamefont {Kuang}}, \ and\ \bibinfo
  {author} {\bibfnamefont {Q.}~\bibnamefont {Lu}},\ }\href {\doibase
  10.1103/PhysRevD.66.014019} {\bibfield  {journal} {\bibinfo  {journal} {Phys.
  Rev. D}\ }\textbf {\bibinfo {volume} {66}},\ \bibinfo {pages} {014019}
  (\bibinfo {year} {2002})}\BibitemShut {NoStop}%
\bibitem [{\citenamefont {Jiang}\ \emph {et~al.}(2010)\citenamefont {Jiang},
  \citenamefont {Zhang}, \citenamefont {Li},\ and\ \citenamefont
  {Wang}}]{Jiang:2009uf}%
  \BibitemOpen
  \bibfield  {author} {\bibinfo {author} {\bibfnamefont {S.-Z.}\ \bibnamefont
  {Jiang}}, \bibinfo {author} {\bibfnamefont {Y.}~\bibnamefont {Zhang}},
  \bibinfo {author} {\bibfnamefont {C.}~\bibnamefont {Li}}, \ and\ \bibinfo
  {author} {\bibfnamefont {Q.}~\bibnamefont {Wang}},\ }\href {\doibase
  10.1103/PhysRevD.81.014001} {\bibfield  {journal} {\bibinfo  {journal} {Phys.
  Rev. D}\ }\textbf {\bibinfo {volume} {81}},\ \bibinfo {pages} {014001}
  (\bibinfo {year} {2010})}\BibitemShut {NoStop}%
\bibitem [{\citenamefont {Jiang}\ and\ \citenamefont
  {Wang}(2010)}]{Jiang:2010wa}%
  \BibitemOpen
  \bibfield  {author} {\bibinfo {author} {\bibfnamefont {S.-Z.}\ \bibnamefont
  {Jiang}}\ and\ \bibinfo {author} {\bibfnamefont {Q.}~\bibnamefont {Wang}},\
  }\href {\doibase 10.1103/PhysRevD.81.094037} {\bibfield  {journal} {\bibinfo
  {journal} {Phys. Rev. D}\ }\textbf {\bibinfo {volume} {81}},\ \bibinfo
  {pages} {094037} (\bibinfo {year} {2010})}\BibitemShut {NoStop}%
\bibitem [{\citenamefont {Jiang}\ \emph {et~al.}(2013)\citenamefont {Jiang},
  \citenamefont {Wang},\ and\ \citenamefont {Zhang}}]{Jiang:2012ir}%
  \BibitemOpen
  \bibfield  {author} {\bibinfo {author} {\bibfnamefont {S.-Z.}\ \bibnamefont
  {Jiang}}, \bibinfo {author} {\bibfnamefont {Q.}~\bibnamefont {Wang}}, \ and\
  \bibinfo {author} {\bibfnamefont {Y.}~\bibnamefont {Zhang}},\ }\href
  {\doibase 10.1103/PhysRevD.87.094014} {\bibfield  {journal} {\bibinfo
  {journal} {Phys. Rev. D}\ }\textbf {\bibinfo {volume} {87}},\ \bibinfo
  {pages} {094014} (\bibinfo {year} {2013})}\BibitemShut {NoStop}%
\bibitem [{\citenamefont {Jiang}\ \emph {et~al.}(2015)\citenamefont {Jiang},
  \citenamefont {Wei}, \citenamefont {Chen},\ and\ \citenamefont
  {Wang}}]{Jiang:2015dba}%
  \BibitemOpen
  \bibfield  {author} {\bibinfo {author} {\bibfnamefont {S.-Z.}\ \bibnamefont
  {Jiang}}, \bibinfo {author} {\bibfnamefont {Z.-L.}\ \bibnamefont {Wei}},
  \bibinfo {author} {\bibfnamefont {Q.-S.}\ \bibnamefont {Chen}}, \ and\
  \bibinfo {author} {\bibfnamefont {Q.}~\bibnamefont {Wang}},\ }\href {\doibase
  10.1103/PhysRevD.92.025014} {\bibfield  {journal} {\bibinfo  {journal} {Phys.
  Rev. D}\ }\textbf {\bibinfo {volume} {92}},\ \bibinfo {pages} {025014}
  (\bibinfo {year} {2015})}\BibitemShut {NoStop}%
\bibitem [{\citenamefont {Wang}\ and\ \citenamefont
  {Wang}(2000)}]{Wang:2000mu}%
  \BibitemOpen
  \bibfield  {author} {\bibinfo {author} {\bibfnamefont {X.-L.}\ \bibnamefont
  {Wang}}\ and\ \bibinfo {author} {\bibfnamefont {Q.}~\bibnamefont {Wang}},\
  }\href@noop {} {\bibfield  {journal} {\bibinfo  {journal} {Commun. Theor.
  Phys.}\ }\textbf {\bibinfo {volume} {34}},\ \bibinfo {pages} {519} (\bibinfo
  {year} {2000})}\BibitemShut {NoStop}%
\bibitem [{\citenamefont {Wang}\ \emph
  {et~al.}(2000{\natexlab{b}})\citenamefont {Wang}, \citenamefont {Wang},\ and\
  \citenamefont {Wang}}]{Wang:2000mg}%
  \BibitemOpen
  \bibfield  {author} {\bibinfo {author} {\bibfnamefont {X.-L.}\ \bibnamefont
  {Wang}}, \bibinfo {author} {\bibfnamefont {Z.-M.}\ \bibnamefont {Wang}}, \
  and\ \bibinfo {author} {\bibfnamefont {Q.}~\bibnamefont {Wang}},\ }\href@noop
  {} {\bibfield  {journal} {\bibinfo  {journal} {Commun. Theor. Phys.}\
  }\textbf {\bibinfo {volume} {34}},\ \bibinfo {pages} {683} (\bibinfo {year}
  {2000}{\natexlab{b}})}\BibitemShut {NoStop}%
\bibitem [{\citenamefont {Ren}\ \emph {et~al.}(2017)\citenamefont {Ren},
  \citenamefont {Fu},\ and\ \citenamefont {Wang}}]{Ren:2017bhd}%
  \BibitemOpen
  \bibfield  {author} {\bibinfo {author} {\bibfnamefont {K.}~\bibnamefont
  {Ren}}, \bibinfo {author} {\bibfnamefont {H.-F.}\ \bibnamefont {Fu}}, \ and\
  \bibinfo {author} {\bibfnamefont {Q.}~\bibnamefont {Wang}},\ }\href {\doibase
  10.1103/PhysRevD.95.074012} {\bibfield  {journal} {\bibinfo  {journal} {Phys.
  Rev. D}\ }\textbf {\bibinfo {volume} {95}},\ \bibinfo {pages} {074012}
  (\bibinfo {year} {2017})}\BibitemShut {NoStop}%
\bibitem [{\citenamefont {Chen}\ \emph {et~al.}(2020)\citenamefont {Chen},
  \citenamefont {Fu}, \citenamefont {Ma},\ and\ \citenamefont
  {Wang}}]{Chen2020jiq}%
  \BibitemOpen
  \bibfield  {author} {\bibinfo {author} {\bibfnamefont {Q.-S.}\ \bibnamefont
  {Chen}}, \bibinfo {author} {\bibfnamefont {H.-F.}\ \bibnamefont {Fu}},
  \bibinfo {author} {\bibfnamefont {Y.-L.}\ \bibnamefont {Ma}}, \ and\ \bibinfo
  {author} {\bibfnamefont {Q.}~\bibnamefont {Wang}},\ }\href {\doibase
  10.1103/PhysRevD.102.034034} {\bibfield  {journal} {\bibinfo  {journal}
  {Phys. Rev. D}\ }\textbf {\bibinfo {volume} {102}},\ \bibinfo {pages}
  {034034} (\bibinfo {year} {2020})},\ \Eprint
  {http://arxiv.org/abs/2001.06418} {arXiv:2001.06418 [hep-ph]} \BibitemShut
  {NoStop}%
\bibitem [{\citenamefont {Wise}(1992)}]{Wise:1992hn}%
  \BibitemOpen
  \bibfield  {author} {\bibinfo {author} {\bibfnamefont {M.~B.}\ \bibnamefont
  {Wise}},\ }\href {\doibase 10.1103/PhysRevD.45.R2188} {\bibfield  {journal}
  {\bibinfo  {journal} {Phys. Rev. D}\ }\textbf {\bibinfo {volume} {45}},\
  \bibinfo {pages} {R2188} (\bibinfo {year} {1992})}\BibitemShut {NoStop}%
\bibitem [{\citenamefont {Yan}\ \emph {et~al.}(1992)\citenamefont {Yan},
  \citenamefont {Cheng}, \citenamefont {Cheung}, \citenamefont {Lin},
  \citenamefont {Lin},\ and\ \citenamefont {Yu}}]{Yan:1992gz}%
  \BibitemOpen
  \bibfield  {author} {\bibinfo {author} {\bibfnamefont {T.-M.}\ \bibnamefont
  {Yan}}, \bibinfo {author} {\bibfnamefont {H.-Y.}\ \bibnamefont {Cheng}},
  \bibinfo {author} {\bibfnamefont {C.-Y.}\ \bibnamefont {Cheung}}, \bibinfo
  {author} {\bibfnamefont {G.-L.}\ \bibnamefont {Lin}}, \bibinfo {author}
  {\bibfnamefont {Y.}~\bibnamefont {Lin}}, \ and\ \bibinfo {author}
  {\bibfnamefont {H.-L.}\ \bibnamefont {Yu}},\ }\href {\doibase
  10.1103/PhysRevD.46.1148} {\bibfield  {journal} {\bibinfo  {journal} {Phys.
  Rev. D}\ }\textbf {\bibinfo {volume} {46}},\ \bibinfo {pages} {1148}
  (\bibinfo {year} {1992})},\ \bibinfo {note} {[Erratum: Phys.Rev.D 55, 5851
  (1997)]}\BibitemShut {NoStop}%
\bibitem [{\citenamefont {Nowak}\ \emph {et~al.}(1993)\citenamefont {Nowak},
  \citenamefont {Rho},\ and\ \citenamefont {Zahed}}]{Nowak:1992um}%
  \BibitemOpen
  \bibfield  {author} {\bibinfo {author} {\bibfnamefont {M.~A.}\ \bibnamefont
  {Nowak}}, \bibinfo {author} {\bibfnamefont {M.}~\bibnamefont {Rho}}, \ and\
  \bibinfo {author} {\bibfnamefont {I.}~\bibnamefont {Zahed}},\ }\href
  {\doibase 10.1103/PhysRevD.48.4370} {\bibfield  {journal} {\bibinfo
  {journal} {Phys. Rev. D}\ }\textbf {\bibinfo {volume} {48}},\ \bibinfo
  {pages} {4370} (\bibinfo {year} {1993})}\BibitemShut {NoStop}%
\bibitem [{\citenamefont {Casalbuoni}\ \emph {et~al.}(1993)\citenamefont
  {Casalbuoni}, \citenamefont {Deandrea}, \citenamefont {Di~Bartolomeo},
  \citenamefont {Gatto}, \citenamefont {Feruglio},\ and\ \citenamefont
  {Nardulli}}]{Casalbuoni:1992dx}%
  \BibitemOpen
  \bibfield  {author} {\bibinfo {author} {\bibfnamefont {R.}~\bibnamefont
  {Casalbuoni}}, \bibinfo {author} {\bibfnamefont {A.}~\bibnamefont
  {Deandrea}}, \bibinfo {author} {\bibfnamefont {N.}~\bibnamefont
  {Di~Bartolomeo}}, \bibinfo {author} {\bibfnamefont {R.}~\bibnamefont
  {Gatto}}, \bibinfo {author} {\bibfnamefont {F.}~\bibnamefont {Feruglio}}, \
  and\ \bibinfo {author} {\bibfnamefont {G.}~\bibnamefont {Nardulli}},\ }\href
  {\doibase 10.1016/0370-2693(93)90895-O} {\bibfield  {journal} {\bibinfo
  {journal} {Phys. Lett. B}\ }\textbf {\bibinfo {volume} {299}},\ \bibinfo
  {pages} {139} (\bibinfo {year} {1993})},\ \Eprint
  {http://arxiv.org/abs/hep-ph/9211248} {arXiv:hep-ph/9211248} \BibitemShut
  {NoStop}%
\bibitem [{\citenamefont {Cheng}\ \emph {et~al.}(1994)\citenamefont {Cheng},
  \citenamefont {Cheung}, \citenamefont {Lin}, \citenamefont {Lin},
  \citenamefont {Yan},\ and\ \citenamefont {Yu}}]{Cheng:1993gc}%
  \BibitemOpen
  \bibfield  {author} {\bibinfo {author} {\bibfnamefont {H.-Y.}\ \bibnamefont
  {Cheng}}, \bibinfo {author} {\bibfnamefont {C.-Y.}\ \bibnamefont {Cheung}},
  \bibinfo {author} {\bibfnamefont {G.-L.}\ \bibnamefont {Lin}}, \bibinfo
  {author} {\bibfnamefont {Y.}~\bibnamefont {Lin}}, \bibinfo {author}
  {\bibfnamefont {T.-M.}\ \bibnamefont {Yan}}, \ and\ \bibinfo {author}
  {\bibfnamefont {H.-L.}\ \bibnamefont {Yu}},\ }\href {\doibase
  10.1103/PhysRevD.49.2490} {\bibfield  {journal} {\bibinfo  {journal} {Phys.
  Rev. D}\ }\textbf {\bibinfo {volume} {49}},\ \bibinfo {pages} {2490}
  (\bibinfo {year} {1994})},\ \Eprint {http://arxiv.org/abs/hep-ph/9308283}
  {arXiv:hep-ph/9308283} \BibitemShut {NoStop}%
\bibitem [{\citenamefont {Balk}\ \emph {et~al.}(1994)\citenamefont {Balk},
  \citenamefont {Korner},\ and\ \citenamefont {Pirjol}}]{Balk:1993ev}%
  \BibitemOpen
  \bibfield  {author} {\bibinfo {author} {\bibfnamefont {S.}~\bibnamefont
  {Balk}}, \bibinfo {author} {\bibfnamefont {J.}~\bibnamefont {Korner}}, \ and\
  \bibinfo {author} {\bibfnamefont {D.}~\bibnamefont {Pirjol}},\ }\href
  {\doibase 10.1016/0550-3213(94)90211-9} {\bibfield  {journal} {\bibinfo
  {journal} {Nucl. Phys. B}\ }\textbf {\bibinfo {volume} {428}},\ \bibinfo
  {pages} {499} (\bibinfo {year} {1994})},\ \Eprint
  {http://arxiv.org/abs/hep-ph/9307230} {arXiv:hep-ph/9307230} \BibitemShut
  {NoStop}%
\bibitem [{\citenamefont {Di~Bartolomeo}\ \emph {et~al.}(1995)\citenamefont
  {Di~Bartolomeo}, \citenamefont {Gatto}, \citenamefont {Feruglio},\ and\
  \citenamefont {Nardulli}}]{DiBartolomeo:1994ir}%
  \BibitemOpen
  \bibfield  {author} {\bibinfo {author} {\bibfnamefont {N.}~\bibnamefont
  {Di~Bartolomeo}}, \bibinfo {author} {\bibfnamefont {R.}~\bibnamefont
  {Gatto}}, \bibinfo {author} {\bibfnamefont {F.}~\bibnamefont {Feruglio}}, \
  and\ \bibinfo {author} {\bibfnamefont {G.}~\bibnamefont {Nardulli}},\ }\href
  {\doibase 10.1016/0370-2693(95)00071-R} {\bibfield  {journal} {\bibinfo
  {journal} {Phys. Lett. B}\ }\textbf {\bibinfo {volume} {347}},\ \bibinfo
  {pages} {405} (\bibinfo {year} {1995})},\ \Eprint
  {http://arxiv.org/abs/hep-ph/9411210} {arXiv:hep-ph/9411210} \BibitemShut
  {NoStop}%
\bibitem [{\citenamefont {Cheung}\ and\ \citenamefont
  {Hwang}(2016)}]{Cheung:2015rya}%
  \BibitemOpen
  \bibfield  {author} {\bibinfo {author} {\bibfnamefont {C.-Y.}\ \bibnamefont
  {Cheung}}\ and\ \bibinfo {author} {\bibfnamefont {C.-W.}\ \bibnamefont
  {Hwang}},\ }\href {\doibase 10.1140/epjc/s10052-016-3883-5} {\bibfield
  {journal} {\bibinfo  {journal} {Eur. Phys. J. C}\ }\textbf {\bibinfo {volume}
  {76}},\ \bibinfo {pages} {19} (\bibinfo {year} {2016})},\ \Eprint
  {http://arxiv.org/abs/1508.07686} {arXiv:1508.07686 [hep-ph]} \BibitemShut
  {NoStop}%
\bibitem [{\citenamefont {Alhakami}(2020)}]{Alhakami:2019ait}%
  \BibitemOpen
  \bibfield  {author} {\bibinfo {author} {\bibfnamefont {M.~H.}\ \bibnamefont
  {Alhakami}},\ }\href {\doibase 10.1103/PhysRevD.101.016001} {\bibfield
  {journal} {\bibinfo  {journal} {Phys. Rev. D}\ }\textbf {\bibinfo {volume}
  {101}},\ \bibinfo {pages} {016001} (\bibinfo {year} {2020})},\ \Eprint
  {http://arxiv.org/abs/1910.04409} {arXiv:1910.04409 [hep-ph]} \BibitemShut
  {NoStop}%
\bibitem [{\citenamefont {Manohar}\ and\ \citenamefont
  {Wise}(2000)}]{Manohar:2000dt}%
  \BibitemOpen
  \bibfield  {author} {\bibinfo {author} {\bibfnamefont {A.~V.}\ \bibnamefont
  {Manohar}}\ and\ \bibinfo {author} {\bibfnamefont {M.~B.}\ \bibnamefont
  {Wise}},\ }\href@noop {} {\bibfield  {journal} {\bibinfo  {journal}
  {Cambridge Monogr. Part. Phys. Nucl. Phys., Cosmol.}\ }\textbf {\bibinfo
  {volume} {10}},\ \bibinfo {pages} {1} (\bibinfo {year} {2000})}\BibitemShut
  {NoStop}%
\bibitem [{\citenamefont {Nowak}\ \emph {et~al.}(2004)\citenamefont {Nowak},
  \citenamefont {Praszalowicz}, \citenamefont {Sadzikowski},\ and\
  \citenamefont {Wasiluk}}]{Nowak:2004jg}%
  \BibitemOpen
  \bibfield  {author} {\bibinfo {author} {\bibfnamefont {M.~A.}\ \bibnamefont
  {Nowak}}, \bibinfo {author} {\bibfnamefont {M.}~\bibnamefont {Praszalowicz}},
  \bibinfo {author} {\bibfnamefont {M.}~\bibnamefont {Sadzikowski}}, \ and\
  \bibinfo {author} {\bibfnamefont {J.}~\bibnamefont {Wasiluk}},\ }\href
  {\doibase 10.1103/PhysRevD.70.031503} {\bibfield  {journal} {\bibinfo
  {journal} {Phys. Rev. D}\ }\textbf {\bibinfo {volume} {70}},\ \bibinfo
  {pages} {031503} (\bibinfo {year} {2004})}\BibitemShut {NoStop}%
\bibitem [{\citenamefont {Bardeen}\ and\ \citenamefont
  {Hill}(1994)}]{Bardeen:1993ae}%
  \BibitemOpen
  \bibfield  {author} {\bibinfo {author} {\bibfnamefont {W.~A.}\ \bibnamefont
  {Bardeen}}\ and\ \bibinfo {author} {\bibfnamefont {C.~T.}\ \bibnamefont
  {Hill}},\ }\href {\doibase 10.1103/PhysRevD.49.409} {\bibfield  {journal}
  {\bibinfo  {journal} {Phys. Rev. D}\ }\textbf {\bibinfo {volume} {49}},\
  \bibinfo {pages} {409} (\bibinfo {year} {1994})}\BibitemShut {NoStop}%
\bibitem [{\citenamefont {Zyla}\ \emph {et~al.}(2020)\citenamefont {Zyla} \emph
  {et~al.}}]{ZylaPDG}%
  \BibitemOpen
  \bibfield  {author} {\bibinfo {author} {\bibfnamefont {P.}~\bibnamefont
  {Zyla}} \emph {et~al.} (\bibinfo {collaboration} {Particle Data Group}),\
  }\href@noop {} {\bibfield  {journal} {\bibinfo  {journal} {Prog. Theor. Exp.
  Phys.}\ }\textbf {\bibinfo {volume} {2020}},\ \bibinfo {pages} {083C01}
  (\bibinfo {year} {2020})}\BibitemShut {NoStop}%
\bibitem [{\citenamefont {Pagels}\ and\ \citenamefont
  {Stokar}(1979)}]{Pagels:1979hd}%
  \BibitemOpen
  \bibfield  {author} {\bibinfo {author} {\bibfnamefont {H.}~\bibnamefont
  {Pagels}}\ and\ \bibinfo {author} {\bibfnamefont {S.}~\bibnamefont
  {Stokar}},\ }\href {\doibase 10.1103/PhysRevD.20.2947} {\bibfield  {journal}
  {\bibinfo  {journal} {Phys. Rev. D}\ }\textbf {\bibinfo {volume} {20}},\
  \bibinfo {pages} {2947} (\bibinfo {year} {1979})}\BibitemShut {NoStop}%
\bibitem [{\citenamefont {Dudal}\ \emph {et~al.}(2012)\citenamefont {Dudal},
  \citenamefont {Oliveira},\ and\ \citenamefont
  {Rodriguez-Quintero}}]{Dudal:2012zx}%
  \BibitemOpen
  \bibfield  {author} {\bibinfo {author} {\bibfnamefont {D.}~\bibnamefont
  {Dudal}}, \bibinfo {author} {\bibfnamefont {O.}~\bibnamefont {Oliveira}}, \
  and\ \bibinfo {author} {\bibfnamefont {J.}~\bibnamefont
  {Rodriguez-Quintero}},\ }\href {\doibase 10.1103/PhysRevD.86.105005,
  10.1103/PhysRevD.86.109902} {\bibfield  {journal} {\bibinfo  {journal} {Phys.
  Rev. D}\ }\textbf {\bibinfo {volume} {86}},\ \bibinfo {pages} {105005}
  (\bibinfo {year} {2012})}\BibitemShut {NoStop}%
\bibitem [{\citenamefont {Oliveira}\ \emph {et~al.}(2019)\citenamefont
  {Oliveira}, \citenamefont {de~Paula}, \citenamefont {Frederico},\ and\
  \citenamefont {de~Melo}}]{Oliveira:2018ukh}%
  \BibitemOpen
  \bibfield  {author} {\bibinfo {author} {\bibfnamefont {O.}~\bibnamefont
  {Oliveira}}, \bibinfo {author} {\bibfnamefont {W.}~\bibnamefont {de~Paula}},
  \bibinfo {author} {\bibfnamefont {T.}~\bibnamefont {Frederico}}, \ and\
  \bibinfo {author} {\bibfnamefont {J.}~\bibnamefont {de~Melo}},\ }\href
  {\doibase 10.1140/epjc/s10052-019-6617-7} {\bibfield  {journal} {\bibinfo
  {journal} {Eur. Phys. J. C}\ }\textbf {\bibinfo {volume} {79}},\ \bibinfo
  {pages} {116} (\bibinfo {year} {2019})}\BibitemShut {NoStop}%
\bibitem [{\citenamefont {Fu}\ \emph {et~al.}(2017)\citenamefont {Fu},
  \citenamefont {Wang},\ and\ \citenamefont {Jiang}}]{Fu:2017azw}%
  \BibitemOpen
  \bibfield  {author} {\bibinfo {author} {\bibfnamefont {H.-F.}\ \bibnamefont
  {Fu}}, \bibinfo {author} {\bibfnamefont {Q.}~\bibnamefont {Wang}}, \ and\
  \bibinfo {author} {\bibfnamefont {L.}~\bibnamefont {Jiang}},\ }\href
  {\doibase 10.1103/PhysRevD.96.094023} {\bibfield  {journal} {\bibinfo
  {journal} {Phys. Rev. D}\ }\textbf {\bibinfo {volume} {96}},\ \bibinfo
  {pages} {094023} (\bibinfo {year} {2017})},\ \Eprint
  {http://arxiv.org/abs/1706.03181} {arXiv:1706.03181 [hep-th]} \BibitemShut
  {NoStop}%
\bibitem [{\citenamefont {Tandy}(1997)}]{Tandy:1997qf}%
  \BibitemOpen
  \bibfield  {author} {\bibinfo {author} {\bibfnamefont {P.~C.}\ \bibnamefont
  {Tandy}},\ }\href {\doibase 10.1016/S0146-6410(97)00043-4} {\bibfield
  {journal} {\bibinfo  {journal} {Prog. Part. Nucl. Phys.}\ }\textbf {\bibinfo
  {volume} {39}},\ \bibinfo {pages} {117} (\bibinfo {year} {1997})}\BibitemShut
  {NoStop}%
\end{thebibliography}%

\end{document}